\documentclass[sigconf]{acmart}

\setcopyright{none}
\renewcommand\footnotetextcopyrightpermission[1]{}

\definecolor{boarder}{HTML}{B3B3B3} 
\definecolor{box}{HTML}{FAFAFA}     

\colorlet{framecolor}{boarder}
\colorlet{shadecolor}{box}
\usepackage{multirow}
\usepackage{tabularx}
\usepackage{CJKutf8}
\usepackage{graphicx}   
\usepackage{subcaption} 
\usepackage{framed}
\usepackage{booktabs} 
\usepackage{siunitx} 
\usepackage{array}
\usepackage[utf8]{inputenc}
\usepackage{makecell}
\usepackage{placeins,mydefs}

\AtBeginDocument{%
  }

\renewcommand\acmConference[4]{}
\renewcommand\acmBooktitle[1]{}

\begin{document}
\thispagestyle{empty}
\pagestyle{empty}

\title{Echoes of Norms: Investigating Counterspeech Bots' Influence on Bystanders in Online Communities}

\author{Mengyao Wang}
\affiliation{%
  \institution{Fudan University}
  \city{Shanghai}
  \country{China}}
\email{mengyaowang23@m.fudan.edu.cn}

\author{Shuai Ma}
\affiliation{%
  \institution{Chinese Academy of Sciences}
  \city{Beijing}
  \country{China}}
\email{mashuai@iscas.ac.cn}

\author{Nuo Li}
\affiliation{%
  \institution{Fudan University}
  \city{Shanghai}
  \country{China}}
\email{linuo@fudan.edu.cn}

\author{Peng Zhang}
\authornotemark[1]
\affiliation{%
  \institution{Fudan University}
  \city{Shanghai}
  \country{China}}
\email{zhangpeng_@fudan.edu.cn}

\author{Chenxin Li}
\affiliation{%
  \institution{Fudan University}
  \city{Shanghai}
  \country{China}}
\email{24300740004@m.fudan.edu.cn}

\author{Ning Gu}
\affiliation{%
  \institution{Fudan University}
  \city{Shanghai}
  \country{China}}
\email{ninggu@fudan.edu.cn}

\author{Tun Lu}
\authornote{Corresponding authors.}
\affiliation{%
  \institution{Fudan University}
  \city{Shanghai}
  \country{China}}
\email{lutun@fudan.edu.cn}

\renewcommand{\shortauthors}{Wang et al.}

\begin{abstract}

Counterspeech offers a non-repressive approach to moderate hate speech in online communities. Research has examined how counterspeech chatbots restrain hate speakers and support targets, but their impact on bystanders remains unclear. Therefore, we developed a counterspeech strategy framework and built \textit{Civilbot} for a mixed-method within-subjects study. Bystanders generally viewed Civilbot as credible and normative, though its shallow reasoning limited persuasiveness. Its behavioural effects were subtle: when performing well, it could guide participation or act as a stand-in; when performing poorly, it could discourage bystanders or motivate them to step in. Strategy proved critical: cognitive strategies that appeal to reason, especially when paired with a positive tone, were relatively effective, while mismatch of contexts and strategies could weaken impact. Based on these findings, we offer design insights for mobilizing bystanders and shaping online discourse, highlighting when to intervene and how to do so through reasoning-driven and context-aware strategies.


\end{abstract}

\ccsdesc[500]{Human-centered computing~Empirical studies in collaborative and social computing}

\keywords{Online communities, Hate speech, Counterspeech, Bystanders, Chatbot}

\maketitle
\section{Introduction}
\label{sec1 introduction}
Online communities aspire to foster a diverse, open, and vibrant space for public, yet this ideal is increasingly undermined by the spread of hate speech. Hate speech is commonly defined as targeted, harmful, and weaponized forms of expression against specific groups \cite{ullmannCounterspeechMultidisciplinaryPerspectives2024}. It inflicts profound emotional and psychological harm on its victims, ranging from anxiety and self-blame to suicidal ideation, and can even trigger real-world violence \cite{aleksandricSadnessAngerAnxiety2023, aleksandricUsersBehavioralEmotional2024, bonaldiNLPCounterspeechHate2024}. A particularly troubling aspect is how easily hate speech spreads: when people are exposed to malicious or antisocial comments, they become more likely to produce similar negativity \cite{aleksandricUsersBehavioralEmotional2024, chengAnyoneCanBecome2017}, even if they are usually not highly aggressive \cite{saveskiStructureToxicConversations2021}. Within communities, hate speech propagates imitation, perpetuates prejudice, and reinforces stereotypes \cite{ullmannCounterspeechMultidisciplinaryPerspectives2024}, ultimately fuelling stigmatization \cite{linkConceptualizingStigma2001} and structural injustice \cite{cepollaroCounterspeech2023}, which can further divide society. Traditional content moderation relies on restrictive interventions imposed by platforms to counteract hate speech. However, such methods often risk the over-removal of empowering discourse, forcing platforms to adopt cautious and conservative classification thresholds, which in turn further overlook implicit forms of hate speech \cite{hartmannLostModerationHow2025}. Moreover, users can easily migrate to platforms with looser moderation policies, limiting the effectiveness of isolated interventions \cite{ullmannCounterspeechMultidisciplinaryPerspectives2024}. In light of the limitations of heavy-handed moderation, counterspeech has emerged as a widely recognized alternative—non-repressive, socially grounded, and scalable—that seeks not to suppress but to enrich online discourse.




Counterspeech is typically defined as a direct reply to hateful content \cite{ullmannCounterspeechMultidisciplinaryPerspectives2024, beneschCounteringDangerousSpeech2014, schiebGoverningHateSpeech}, which can take forms such as challenging or condemning the attack \cite{vidgenDetectingEastAsian2020, heRacismVirusAntiasian2022}, expressing solidarity and support \cite{heRacismVirusAntiasian2022}, or even attempting to shift perspectives \cite{qianBenchmarkDatasetLearning2019}. Unlike content moderation, counter speech is rooted in a liberal tradition, emphasizing “speech against speech” rather than “power against speech” \cite{cepollaroCounterspeech2023, langtonBlockingCounterspeech2018}. Its potential lies not only in addressing individual instances of hate but also in influencing broader social norms that frame hate as unacceptable, thereby altering the overall tone of community dialogue \cite{bilewiczArtificialIntelligenceHate2021, berryDiscussionQualityDiffuses2017}. Within such a mechanism, online spaces need not serve as arenas for the spread of hostility but can evolve into fairer and safer environments for communication \cite{ullmannCounterspeechMultidisciplinaryPerspectives2024, sasseBreakingSilenceInvestigating2023}. Moreover, counterspeech may generate a “positive contagion effect”: expressions that denounce incivility or encourage respectful dialogue can themselves inspire similar responses, creating cycles of positive imitation and role modelling \cite{molinaRoleCivilityMetacommunication2018, seeringShapingProAntiSocial2017}.

In recent years, the rapid advancement of generative AI has provided a technical foundation for the automation of counterspeech. A growing body of work has constructed hate-counterspeech corpora \cite{albanyanFindingAuthenticCounterhate2023, halimWokeGPTImprovingCounterspeech2023a, mathewAnalyzingHateCounter2018, yuHateSpeechCounter2022, zhengWhatMakesGood2023b}, trained classification and generation models \cite{mathewAnalyzingHateCounter2018, guptaCounterspeechesMySleeve2023, sahaConsolidatingStrategiesCountering2024, zhuGeneratePruneSelect2021, chungCONANCOunterNArratives2019, benniePANDAPairedAntihate2025}, and introduced enhancement strategies such as incorporating background knowledge \cite{albanyanFindingAuthenticCounterhate2023, wilkFactbasedCounterNarrative2025}, contextual information \cite{barGenerativeAIMay2024, cimaContextualizedCounterspeechStrategies2025a, dogancGenericPersonalizedInvestigating2023}, and complex argumentative structures \cite{guptaCounterspeechesMySleeve2023, sahaConsolidatingStrategiesCountering2024, hangartnerEmpathybasedCounterspeechCan2021}, etc. These approaches have enabled counterspeech chatbots to achieve more persuasive argumentation and contextually appropriate responses in online community settings, amplifying potential social impact. However, existing research on counterspeech chatbots has primarily focused on curbing hate speakers \cite{bilewiczArtificialIntelligenceHate2021, barGenerativeAIMay2024, hangartnerEmpathybasedCounterspeechCan2021} or providing support for targets of hate \cite{pingCounterExploringMotivations2024, pingPerceivingCounteringHate2025}, while paying far less attention to bystanders—arguably the most crucial group in terms of both size and influence. Bystanders are not neutral observers: they often lean toward those who oppose hate speech, forming a “silent majority” that substantially shapes the community’s prevailing attitudes and social norms \cite{ullmannCounterspeechMultidisciplinaryPerspectives2024}. They may become potential counterspeakers, yet their silence can also be interpreted as implicit tolerance of hate. Prior research suggests that public counterspeech can disrupt the perception that “most people tolerate hate”, thereby weakening the spiral of silence \cite{noelle-neumannSpiralSilenceTheory1974}, and may also trigger herd effects \cite{baddeleyHerdingSocialInfluence2010} that motivate more users to speak out. However, it remains unclear whether counterspeech chatbots can exert comparable social influence, as systematic empirical evidence is still limited.

To this end, we focus on the social influence of counterspeech chatbots on bystanders in online communities and how these effects ripple into the broader normative climate. We therefore pose the following research questions:

\begin{itemize}
    \item RQ1: To what extent do bystanders endorse the chatbot’s counterspeech and show changes in their behavioral tendencies (e.g., perceived reason strength, credibility, and confidence in countering)?
    \item RQ2: How do different types of chatbot' counterspeech shape bystanders’ endorsement of the chatbot’s responses and change of their behavioral tendencies?
\end{itemize}

To establish a structured foundation for studying counterspeech, we developed a unified framework along three dimensions: sentence type (question and non-question), tone (positive and negative), and strategic intent, which includes cognitive strategies (e.g., highlighting truth) and affective strategies (e.g., denouncing hate speakers). The combination of dimensions produced eight distinct counterspeech strategies. Based on this framework, we built Civilbot, a prototype chatbot that generates context-aware counterspeech across these strategies. We then conducted a mixed-methods within-subject experiment with participants recruited from online communities who were interested in sensitive topics, generally silent in public discussions, yet opposed to hate speech. Each participant chose eight topics and, for each, read a hate-speech post, observed a counterspeech reply, and completed pre- and post-exposure attitude measures. Sessions were randomly assigned to the eight strategies, so everyone experienced all strategies. We assessed perceived counterspeech quality (e.g., convincing and strong reasons), subjective acceptance (e.g., credibility, importance, overall agreement), and behavioural tendencies (e.g., confidence in countering, willingness to participate), and complemented these metrics with semi-structured interviews exploring bystanders' detailed perceptions of chatbot roles, and potential effects on community norms, etc.

Our findings show that Civilbot’s counterspeech shapes bystanders’ perceptions of both the counterspeech itself and the bot, and also influences the overall climate of online communities—even affecting subtle behavioural tendencies. Civilbot is generally viewed as credible and as signalling community norms, though its shallow reasoning constrains its persuasiveness. Behaviourally, its impact is subtle and sometimes mixed: it can guide and encourage bystanders, substitute for their own response, dampen participation when it performs poorly, or prompt users to step in out of frustration with its limitations. Beyond persuasion, it also contributes to the broader community atmosphere, such as by helping to cool down emotional intensity or provide additional information for reflection. Strategy proves decisive: cognitive strategies are typically more effective than affective ones; tone may influence behavioural tendencies but relies on specific contexts; and sentence forms, when paired with other strategies, can either stimulate reflection or trigger resistance. These findings offer design insights for Civilbot to engage bystanders and help mediate a hostile community climate: Civilbot must determine when to intervene and why its intervention is needed, and it must also decide how to intervene through reasoning-driven motivation, information-enabled argument, context-adaptive strategy, and extending modalities beyond text.

Our work makes the following contributions:

\begin{itemize}
    \item To the best of our knowledge, this is the first study that examines the mechanisms through which counterspeech chatbots influence bystanders in online hate incidents.
    \item We develop an experimental framework for counterspeech that incorporates three key dimensions (sentence type, tone, and strategic intent). Based on this framework, we build a chatbot prototype (Civilbot) to empirically examine its influence on bystanders' attitudes and behavioural intentions.
    \item We provide design insights for future counterspeech chatbots that explicitly consider the role of bystanders in shaping responses to hate speech within online communities.
\end{itemize}


\section{Related Work}
\label{sec2 related work}
\subsection{Origins and Functions of Counterspeech}
The idea of mitigating the harms of hate speech through counterspeech originates from debates on freedom of expression. Rather than silencing people, this perspective argues that harms can be reduced by responding in constructive ways \cite{cepollaroCounterspeech2023}. In contrast, coercive suppression often entails moral costs: it may further restrict the expressive freedom of marginalized groups, thereby reinforcing exclusion and resentment \cite{lepoutreDemocraticSpeechDivided2021}. Algorithmic moderation is now the dominant response, but platform-level opacity, over-moderation, and exclusion from decision-making have raised concerns about fairness and legitimacy \cite{hartmannLostModerationHow2025, munCounterspeakersPerspectivesUnveiling2024}. Even human moderation often functions as an ex-post control \cite{schlugerProactiveModerationOnline2022}. Against this backdrop, counterspeech has re-emerged as an alternative that promotes social justice: the way to counter falsehood is not suppression but exposure, debate, and persuasion, allowing truth to prevail in open contestation \cite{langtonBlockingCounterspeech2018, bromellCounterSpeechEveryonesResponsibility2022}.

Subsequent research has examined how counterspeech affects online social interactions. It can trigger contagion effects and act as implicit cues of community norms: for example, polite responses increase willingness to engage \cite{hanPlayingNiceModeling2015}, while metacommunication (calling out incivility) promotes civility \cite{molinaRoleCivilityMetacommunication2018}. Emotional contagion and imitation also play a role, with users adapting to the affective tone and behaviour of others \cite{kramerExperimentalEvidenceMassivescale2014, seeringShapingProAntiSocial2017}. These dynamics alter perceptions of norms—both descriptive (what is common) and injunctive (what is appropriate)—and thereby influence whether hate appears tolerated or unwelcome \cite{bilewiczArtificialIntelligenceHate2021, sasseBreakingSilenceInvestigating2023}. 

The social impact of counterspeech is shaped not only by its content but also by the identity of the counterspeaker. Counterspeech can originate from targets of hate, bystanders, or non-targeted users, and may be delivered by ordinary members or authority figures \cite{cepollaroCounterspeech2023}. Interventions from influential or high-status members are more likely to be adopted, while “outsiders” without established identities exert weaker impact \cite{beneschCounteringDangerousSpeech2014, seeringShapingProAntiSocial2017}. Moreover, factors such as race or follower count can further moderate influence \cite{mungerTweetmentEffectsTweeted2017}.

In sum, prior work highlights the promise of counterspeech in shaping community norms through mechanisms such as contagion, emotion, and identity influence. Inspired by this perspective, AI chatbots may serve as consistent and scalable counterspeakers, activating normative cues and emotional dynamics in ways that differs from human interventions.

\subsection{Strategies and Automated Generation of Counterspeech}
As scholarly attention to counterspeech has grown, researchers have proposed diverse taxonomies of strategies. Early work on Twitter identified eight non-exclusive forms, including factual correction, highlighting contradictions, warning of consequences, expressing identification, denouncing hate, using media, humor, or particular tones \cite{ruthsCounterspeechTwitterField2016}. Later studies refined these categories by distinguishing positive versus hostile tones \cite{mathewThouShaltNot2019}, grouping responses into informative, denouncing, questioning, positive, and humorous intents \cite{guptaCounterspeechesMySleeve2023}, or emphasizing empathetic, consequence-warning, and polite formulations \cite{barGenerativeAIMay2024, sahaCrowdCounterBenchmarkTypespecific2024}. Other work has drawn on argument structures, speech acts, and psychological mechanisms such as normative influence and empathy induction \cite{sahaConsolidatingStrategiesCountering2024, dogancGenericPersonalizedInvestigating2023}. Surveys summarize these efforts along broader axes such as active versus passive, positive versus negative style, and responses to explicit versus implicit hate \cite{cepollaroCounterspeech2023}.

Parallel to this conceptual work, rapid advances in AI have enabled the automated generation of counterspeech. Researchers have constructed a range of datasets—from Twitter and online articles \cite{mathewAnalyzingHateCounter2018, albanyanFindingAuthenticCounterhate2023} to large domain-specific corpora such as WokeCorpus \cite{halimWokeGPTImprovingCounterspeech2023a} and Reddit-based annotations \cite{yuHateSpeechCounter2022}. These resources support models that generate counterspeech \cite{guptaCounterspeechesMySleeve2023, sahaConsolidatingStrategiesCountering2024, benniePANDAPairedAntihate2025}, experiment with large language models \cite{halimWokeGPTImprovingCounterspeech2023a, wilkFactbasedCounterNarrative2025}, and incorporate contextual information such as argumentation, psychology, or personalization \cite{barGenerativeAIMay2024, cimaContextualizedCounterspeechStrategies2025a}. At the same time, HCI studies have explored human–AI co-creation frameworks and design guidelines for AI-assisted counterspeech \cite{dingCounterQuillInvestigatingPotential2025, munCounterspeakersPerspectivesUnveiling2024}.

Overall, prior work has developed diverse but fragmented taxonomies of strategies, and demonstrated the feasibility of automated generation. Yet the lack of a unifying framework limits comparability across approaches. Building on these foundations, our study introduces a structured strategy framework and empirically examines how chatbot-mediated counterspeech influences online communities.

\subsection{Social Impact of Chatbot-Generated Counterspeech}
Existing research on the social impact of counterspeech primarily focuses on two directions: constraining hate speakers and supporting targets of hate. Studies on hate speakers examine both reflections on past hateful behaviours—such as deleting hateful comments \cite{barGenerativeAIMay2024, hangartnerEmpathybasedCounterspeechCan2021}—and potential changes in future behaviours, including shifts in the toxicity of expressed opinions \cite{barGenerativeAIMay2024} or reductions in hate speech and aggression \cite{bilewiczArtificialIntelligenceHate2021, hangartnerEmpathybasedCounterspeechCan2021}. These findings, derived from data analysis and controlled experiments, have also informed design implications for counterspeech chatbot to affect hate speakers. Research on targets of hate, in contrast, investigates how this identity influences their attitudes and behaviours of countering hate. For instance, exposure to online hate can become a key motivator for sustained participation in counterspeech \cite{pingCounterExploringMotivations2024}. Moreover, users whose identities closely align with those of the targets are more likely to perceive counterspeech as a feasible response and to actively engage in it \cite{pingPerceivingCounteringHate2025}.

In contrast, HCI research emphasizes the crucial role of bystanders—the “silent majority” of a community who shape mainstream norms \cite{ullmannCounterspeechMultidisciplinaryPerspectives2024}. The value of interventions stems from their ability to trigger the observer effect, prompting users to align self-expression with community expectations \cite{sahaObserverEffectSocial2024}. Even when targeting violators, these visible interventions function as a “deterrence” mechanism \cite{jhaverBystandersOnlineModeration2024}. Bystanders learn descriptive norms by observing others' behaviours and their consequences, and learn injunctive norms through the explicit behavioural guidelines presented in the interventions \cite{sasseBreakingSilenceInvestigating2023}. Studies have designed interventions to stimulate prosocial bystander behaviour. For instance, interface designs simulating “under observation” contexts, e.g., displaying audience size metrics \cite{difranzoUpstandingDesignBystander2018} or notifying relevant groups that bystanders have viewed cyberbullying content \cite{taylorAccountabilityEmpathyDesign2019}, have been shown to heighten accountability. Similarly, improving moderation transparency via public post-removal explanations helps bystanders understand acceptability boundaries, fostering norm-aware contributions \cite{jhaverDoesTransparencyModeration2019} and significantly boosting engagement levels \cite{jhaverBystandersOnlineModeration2024}. As a similarly information-rich intervention, counterspeech has garnered increasing attention regarding its impact on bystanders. While strategies like condemnation or distraction may yield limited behavioural intent changes \cite{jiaTacklingHateSpeech2025}, perspective-focused strategies may reduce the spread and amplification of hate speech on platforms \cite{gennaroCounterspeechEncouragingUsers2025}. Furthermore, effective counterspeech not only empowers bystanders to actively challenge hate but also reduces the likelihood of future hateful content creation by both bystanders and hate speakers \cite{cyprisEffectivenessCounterspeechMitigating2024}.

While prior work has shown how chatbot-generated counterspeech can restrain hate speakers and support targets, its effects on bystanders remain largely unexplored—even though the value of bystander intervention has been extensively validated in HCI research. Investigating how chatbot-mediated counterspeech interacts with this silent majority is therefore crucial, both to fill a theoretical gap and to derive design insights for chatbots that promote anti-hate and constructive online discourse.


\section{Constructing a Framework of Counterspeech Strategies}
\label{sec3 framework}

\subsection{Methodology of constructing strategy framework}
To synthesize commonly used counterspeech strategies from prior research, we conducted a literature review on Google Scholar focusing on studies related to hate speech interventions, counterspeech generation, and strategy typologies. Building upon existing labels of counterspeech strategies reported in the literature \cite{ruthsCounterspeechTwitterField2016, mathewThouShaltNot2019, guptaCounterspeechesMySleeve2023, sahaCrowdCounterBenchmarkTypespecific2024, barGenerativeAIMay2024, sahaConsolidatingStrategiesCountering2024, dogancGenericPersonalizedInvestigating2023, cepollaroCounterspeech2023} and guided by inductive coding methods \cite{thomasGeneralInductiveApproach2003}, we developed a unified framework. Initially, one author reviewed the literature and extracted text segments pertaining to counterspeech strategies. Two authors then independently identified and annotated preliminary strategy types based on these excerpts. Ambiguities and disagreements were resolved through cross-validation and group discussions, which included semantic refinements to minimize overlap and redundancy among categories. For example, the label "question" appeared in prior studies with different meanings. In some cases, it referred to counter question form \cite{guptaCounterspeechesMySleeve2023, sahaConsolidatingStrategiesCountering2024, chungCONANCOunterNArratives2019, mathewThouShaltNot2019, chungUnderstandingCounterspeechOnline2023, sahaCrowdCounterBenchmarkTypespecific2024}; in others, it denoted challenges to the credibility of information sources or claims underlying hate speech \cite{sahaCrowdCounterBenchmarkTypespecific2024}. To reduce confusion, we treated the first as "question" and the second as part of the "pointing out hypocrisy or contradictions" sub-strategy. This process resulted in the identification of 20 initial sub-strategies.

We then inductively consolidated all initial sub-strategies into eight mutually exclusive categories: \textit{question}, \textit{non-question}, \textit{positive tone}, \textit{negative tone}, \textit{rebutting falsehoods}, \textit{highlighting truth}, \textit{denouncing hate speakers}, and \textit{supporting targets of hate}. Guided by the Elaboration Likelihood Model (ELM) \cite{pettyElaborationLikelihoodModel1986}, the latter four strategies were further synthesized into two overarching types: \textit{cognitive strategy} and \textit{affective strategy}. The final classification framework comprises three dimensions—sentence type (question vs. non-question), tone (positive vs. negative), and strategic intent (cognitive vs. affective influence). The resulting framework is both theoretically grounded and practically oriented, facilitating structured experimental design.

\subsection{Classification Dimensions}

\begin{table*}[t]
  \caption{Framework of Counterspeech Strategies}
  \label{tab:framework}
  \centering
  \renewcommand{\arraystretch}{1.5}
  \footnotesize

  \makebox[\textwidth][c]{%
    \begin{tabular}{
      @{}
      >{\centering\arraybackslash}m{2.6cm}
      >{\centering\arraybackslash}m{3.8cm}
      >{\centering\arraybackslash}m{3.8cm}
      >{\centering\arraybackslash}m{3.8cm}
      @{}
    }
      \toprule
      \normalsize \textbf{Dimension} &
      \normalsize \textbf{Sub-dimension} &
      \normalsize \textbf{Methods \& Explanation} &
      \normalsize \textbf{Example} \\
      \midrule

      \multirow{4}{*}{\textbf{\small Sentence Type}} 
      & \textbf{Question} 
      & Counterspeech framed as questions to challenge hate speakers, prompt reflection, or expose contradictions. 
      & ``If this stereotype were true, how do you explain the many successful Uyghur entrepreneurs in tech and finance?'' \\
      \cline{2-4}

      & \textbf{Non-question} 
      & Includes statements, imperatives, exclamations; directly presents counter-arguments without question framing. 
      & ``Your claim ignores census data showing that crime rates are not higher in immigrant communities.'' \\
      \hline

      \multirow{3}{*}{\textbf{\small Tone}}
      & \textbf{Positive}  
      & Friendly, cooperative, polite, or constructive tone; designed to reduce defensiveness and encourage dialogue. 
      & ``I see why you might think that, but here's another study that tells a different story.'' \\
      \cline{2-4}

      & \textbf{Negative (Sarcasm / Hostile)} 
      & Confrontational, mocking, or hostile tone; can provoke shame or resistance but may discourage repetition. 
      & ``Right, because decades of peer-reviewed research are clearly just made up for fun.'' \\
      \hline

      \multirow{9}{*}{\textbf{\small Strategic Intent}}
      & \textbf{Cognitive strategy} 
      & Aims to change beliefs through reasoning, evidence, and fact-checking, encouraging rational evaluation. 
        [\textbf{Rebutting Falsehoods}] Identifying and refuting false claims, contradictions, or unreliable sources. 
        [\textbf{Highlighting Truth}] Presenting accurate facts, verified evidence, or pointing to credible sources; sometimes warning of real consequences. 
      & ``You said immigrants `don't pay taxes,' but IRS data (2022) shows immigrant households contribute over \$330 billion annually in taxes.'' / 
        ``WHO reports confirm vaccines save 4--5 million lives every year. Spreading misinformation only increases public health risks.'' \\
      \cline{2-4}

      & \textbf{Affective strategy} 
      & Aims to trigger emotional reactions, such as shame (toward hate speakers) or empathy (toward targets). 
        [\textbf{Denouncing Hate Speakers}] Labelling the statement as hateful, biased, or harmful to evoke accountability. 
        [\textbf{Supporting Targets}] Expressing solidarity, empathy, or defence of the targeted group to humanize them and restore dignity.
      & ``This remark is racist and fuels dangerous stereotypes that have led to violence offline.'' / 
        ``I stand with Muslim women who choose how they dress—their voices matter more than your prejudice.'' \\
      \bottomrule
    \end{tabular}
  }
\end{table*}

The framework comprises three dimensions: Sentence Type, Tone, and Strategic Intent.

\subsubsection{Sentence Type}
This dimension distinguishes between questions and non-questions. Questions are highlighted as a specific strategy that formulates counterspeech in interrogative form to challenge hate speakers, including approaches variously labelled as counter questions or questioning \cite{guptaCounterspeechesMySleeve2023, sahaConsolidatingStrategiesCountering2024, chungCONANCOunterNArratives2019, mathewThouShaltNot2019, chungUnderstandingCounterspeechOnline2023, sahaCrowdCounterBenchmarkTypespecific2024}. Given that "question" is frequently identified as a distinct strategy type in counterspeech research, and that studies underscore its role in fostering critical thinking \cite{danryDontJustTell2023b}, we consider it necessary to include sentence type as an independent dimension. All other utterances—such as statements, imperatives, or exclamations—are classified as non-questions.

\subsubsection{Tone}
We conceptualize tone in counterspeech as a spectrum but, for experimental controllability, adopt a binary classification of positive versus negative tone. Positive tone encompasses cooperative and friendly expressions, aligning with prior work on perceived interaction quality with conversational agents \cite{sportelliLetsMakeDifference2025a}. Examples include polite, empathetic, or detoxified responses \cite{guptaCounterspeechesMySleeve2023, sahaConsolidatingStrategiesCountering2024, chungCONANCOunterNArratives2019, mathewThouShaltNot2019, chungUnderstandingCounterspeechOnline2023, qianBenchmarkDatasetLearning2019a, sahaCounterGeDiControllableApproach2022}. Negative tone, by contrast, involves more confrontational or critical expressions, often manifested through sarcasm or humour \cite{guptaCounterspeechesMySleeve2023, chungCONANCOunterNArratives2019, hangartnerEmpathybasedCounterspeechCan2021, ruthsCounterspeechTwitterField2016, chungUnderstandingCounterspeechOnline2023, sahaCrowdCounterBenchmarkTypespecific2024}, whose intensity may vary depending on the utterance.

\subsubsection{Strategic Intent}
The intent dimension is divided into cognitive and affective strategies. This classification builds on inductive synthesis of existing strategy typologies and is informed by the ELM \cite{pettyElaborationLikelihoodModel1986}, which distinguishes cognitive, affective, and behavioural persuasion routes. Since behavioural persuasion in ELM refers to effects triggered by individuals' own actions—which falls outside the scope of this study and is not reflected in current strategy types—we exclude it here.

Cognitive strategies aim to shift bystanders' cognition and can be further divided into rebutting falsehoods and highlighting truth \cite{cepollaroCounterspeech2023}. Rebutting falsehoods involves exposing false, contradictory, or hypocritical elements in hate speech \cite{sahaConsolidatingStrategiesCountering2024, ruthsCounterspeechTwitterField2016, chungUnderstandingCounterspeechOnline2023, sahaCrowdCounterBenchmarkTypespecific2024, vidgenDetectingEastAsian2020}, such as rejecting abusive premises \cite{vidgenDetectingEastAsian2020} or questioning information sources and underlying claims \cite{sahaCrowdCounterBenchmarkTypespecific2024}. Saha et al. further extended this category by incorporating Walton's argumentation schemes—Means for Goal, Goal from Means, Source Knowledge, Source Authority, and Rule or Principle \cite{sahaConsolidatingStrategiesCountering2024}—to enrich argumentative methods. Highlighting truth seeks to change cognition by presenting facts \cite{guptaCounterspeechesMySleeve2023, sahaConsolidatingStrategiesCountering2024, ruthsCounterspeechTwitterField2016, chungUnderstandingCounterspeechOnline2023} or citing arguments from online sources \cite{albanyanFindingAuthenticCounterhate2023}. Recommendations to conduct additional verification such as doing more research \cite{qianBenchmarkDatasetLearning2019a} also fall into this category. Additionally, warning about potential online or offline consequences of hate speech \cite{sahaConsolidatingStrategiesCountering2024, barGenerativeAIMay2024, hangartnerEmpathybasedCounterspeechCan2021, ruthsCounterspeechTwitterField2016, sahaCrowdCounterBenchmarkTypespecific2024} is included, as such warnings supplement information rather than directly refute hate.

Affective strategies aim to trigger change by engaging emotions, primarily including denouncing hate speakers and supporting targets of hate. Denouncing hate speakers labels speech as hateful, dangerous, or biased \cite{sahaConsolidatingStrategiesCountering2024, ruthsCounterspeechTwitterField2016, chungUnderstandingCounterspeechOnline2023, vidgenDetectingEastAsian2020}, often by pointing out hate-related keywords \cite{sahaCrowdCounterBenchmarkTypespecific2024, qianBenchmarkDatasetLearning2019a} or warning about their inappropriateness \cite{qianBenchmarkDatasetLearning2019a}, thereby eliciting shame in the hate speaker. Supporting targets of hate includes expressing solidarity with the targeted group \cite{ruthsCounterspeechTwitterField2016, chungUnderstandingCounterspeechOnline2023, sahaCrowdCounterBenchmarkTypespecific2024}, voicing support for specific entities \cite{vidgenDetectingEastAsian2020}, or showing empathy toward the target group \cite{barGenerativeAIMay2024, sahaCrowdCounterBenchmarkTypespecific2024}. The goal here is to make hate speakers aware of the harm caused to others \cite{hangartnerEmpathybasedCounterspeechCan2021}. Compared to denouncing hate speakers, this approach emphasizes empathy rather than shame.

It is worth noting that although some studies propose other strategies such as moral qualities, identity traits, or values \cite{sahaConsolidatingStrategiesCountering2024}, we did not incorporate them into the current framework. Because the literature review indicates that most counterspeech research builds upon, adjusts, and reorganizes earlier classic frameworks \cite{ruthsCounterspeechTwitterField2016} around a relatively stable and reusable set of common strategies. This aligns with the aim of our study: to construct a foundational framework for examining how different strategies influence bystanders. Furthermore, in our experimental design, we explicitly defined the chatbot's identity to avoid confounding anthropomorphic factors. Issues related to chatbot identity, persona, and more complex strategy and expression will be further discussed in Section \ref{sec6 discussion}.

Accordingly, the final framework centers on three binary dimensions—sentence type, tone, and strategic intent—forming a $2 \times 2 \times 2$ taxonomy that yields eight distinct strategy combinations (see Table \ref{tab:framework}). This structure underpins the design of the experimental stimuli. Additionally, each dimension is linked to a set of sub-strategies that function as optional, randomized concrete methods during counterspeech generation. Specifically, within the strategic intent dimension, we only adopt the two primary classifications—cognitive and affective strategies—where further sub-classifications serve as optional, concrete operational methods.

\section{Experimental Design}
\label{sec4 experiment design}
\subsection{Counter Hate Transcript Design}
To support our strategy-based counterspeech experiment, we constructed a dataset of hate speech from the peer-reviewed Chinese bias corpus CDIAL-BIAS DATASET \cite{zhouIdentifyingSocialBias2022}. This dataset is sourced from Zhihu, a widely used Chinese platform known for discussions on diverse social issues, where biased and hateful expressions are prevalent. The corpus covers four major topics of social controversy (gender, race, region, and occupation) and includes multiple subtopics targeting specific groups, offering well-structured content. For example, each entry contains a question and an answer, with the question serving as the context. To extract hate speech from the dataset, we leveraged its existing annotations and selected entries marked as expressing bias or prejudice, which are more likely to contain hateful content. We then applied an operational definition of hate speech—weaponized expressions targeting specific social groups that may cause emotional or psychological harm \cite{ullmannCounterspeechMultidisciplinaryPerspectives2024}—to further refine the data. Using this definition, we designed prompts (see Appendix \ref{app:prompt}) and applied the Qwen-Turbo model to re-screen entries while annotating the targeted groups (e.g., women, immigrants). Two authors then independently reviewed the outputs to determine whether each entry qualified as hate speech and to identify the target groups. Disagreements were resolved through discussion. Ultimately, 27 representative hate speech entries across four categories (gender, race, region, and occupation) were retained as the foundation for strategy-based experimentation. These entries were categorized into several subtypes targeting specific groups (e.g., factory workers), with each subtype comprising multiple distinct questions.


We then used the GPT-5 model and the default decoding settings provided by the API to generate counterspeech texts for the chatbot, adopting an iterative and incremental prompt-engineering process \cite{goversAIDrivenMediationStrategies2024} to ensure that each response adhered to one—and only one—of the eight counterspeech strategies. Building on the strategy framework and definitions introduced in Section \ref{sec3 framework}, we first designed prompts and refined them iteratively. For example, the tone label hostile was revised to sharply critical or emotionally intense to avoid excessively toxic outputs. The final prompt included three components (see Appendix  \ref{app:prompt}): (1) the role definition of the model as a counterspeech generation expert and its goal of producing responses aligned with one of the eight strategy combinations; (2) detailed definitions of each counterspeech dimension and sub-strategy from the perspective of a counterspeaker; and (3) the hate speech text, presented in the format "Hate speech: … Counterspeech: …". To stabilize the generation format, we constructed a small set of eight counterspeech examples that illustrate the expected strategic features. These examples were collaboratively drafted and refined by the authors and remained fixed across all prompts. We adopted a standard few-shot prompting setup to ensure that model outputs conformed consistently to the intended strategy.


To verify that the generated responses correctly implemented the target strategies, two authors independently annotated the initial outputs across three dimensions of the counterspeech framework: sentence type, tone, and strategic intent. As each dimension featured two categories, this created a total of eight distinct target strategies. Inter-rater reliability (IRR) was assessed by calculating Fleiss's Kappa ($\kappa=0.82$) and classification accuracy ($84.42\%$) across the three dimensions in parallel. The resulting $\kappa$ value indicated "almost perfect" agreement \cite{fleissMeasuringNominalScale1971}, which, alongside the high accuracy, established the authors' reliability for subsequent tasks. Responses inconsistent with the target strategy in at least one dimension were flagged for iterative regeneration. For example: “Have you considered how hurtful such comments might be to girls who strive for independence and self-worth? Let’s try to see things from their perspective and respond with understanding rather than blame.” Although this response satisfies the requirements for a positive tone and an affective strategic intent, it mixes question and non-question forms, failing to meet the strategy’s requirement for a purely question sentence type. Therefore, it was flagged for regeneration. This process continued until both authors confirmed complete alignment across all three dimensions, and eight responses for each of the eight strategy types were collected for every hate speech entry. To better simulate a realistic browsing experience in online communities, we also retrieved neutral responses corresponding to the "q"(question) entries in the hate speech dataset from Zhihu. We first collected a broad set of candidate comments via web crawling. Then, two independent researchers then screened the comments to ensure they contained no hateful language, emotional tone, or overt stance-taking. All data were anonymized to protect privacy. Finally, five neutral responses were selected for each question.

\subsection{Bias Mitigation-Counter Hate Transcript Visual Design}
\begin{figure*}[t]
  \centering
  \makebox[\textwidth][c]{%
    \includegraphics[width=0.6\textwidth]{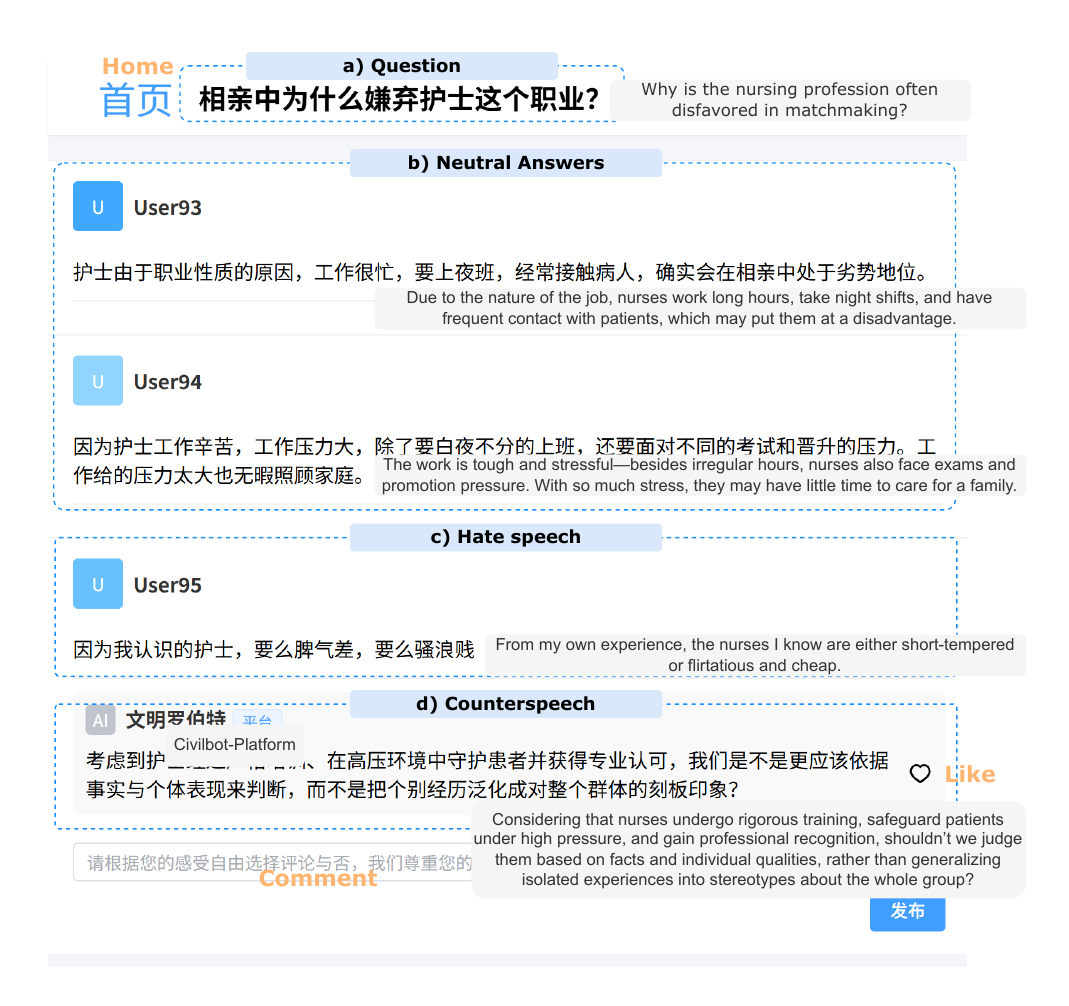}
  }
  \caption{Sample interface of the simulated discussion platform, showing: (a) an excerpted question; (b) neutral answers; (c) a hate-speech post. After participants complete the pre-test questionnaire, the interface displays (d) a counterspeech message. Participants may ``Like'' the counterspeech or post their own comments in response to the hate-speech post. After finishing all interactions, they click ``Home'', complete the post-test questionnaire, and return to the homepage.}
  \label{fig:interface}
\end{figure*}

We selected Zhihu as the design reference for our simulated discussion platform, since our hate speech dataset was originally collected from Zhihu and thus naturally contains the platform's question-answer structure. The system used a Vue \footnote{\url{https://vuejs.org/}} + Element Plus \footnote{\url{https://element-plus.org/}} front end and a lightweight Node.js back end deployed on a remote server. The platform provided two main interfaces: a question-browsing page and a Zhihu-style question page. Following \cite{goversAIDrivenMediationStrategies2024}, we made several design decisions in our Zhihu-style question page visual interface to mitigate potential biases, as illustrated in Figure  \ref{fig:interface}:

\begin{itemize}
    \item Username: In each topic transcript, the hate speakers were assigned different usernames (e.g., User13 for misogynistic hate speech, User12 for hate speech against nurses) to prevent any carry-over effects from prior utterances.
    \item Colour choice: All user avatars were presented in shades of blue, with only tonal variations, to minimize potential bias stemming from avatar imagery or colour-related metaphors.
    \item Removal of interactions: All answer-level interactions, including upvotes, downvotes, and comments, were removed to eliminate implicit cues about community norms that could bias participants.
    \item Answer order: For each question, we displayed a random number (1-5) of neutral answers. The set of neutral responses for each question was fixed across participants to ensure consistency. Neutral answers were always shown first, followed by the hate speech, simulating the experience of unexpectedly encountering hate speech while browsing a topic and providing the necessary contextual buildup for the scenario. To isolate the effect of the counterspeech without potential interference from any subsequent neutral answers, no additional neutral answers were shown afterward, and all measurements were completed immediately after participants viewed the hate speech and the counterspeech.
    \item Timing: After participants encountered the hate speech and completed the first survey measure, counterspeech shows up. This design prevented participants from overlooking counterspeech or being influenced by it prematurely. A second survey measure was then triggered once participants had viewed the counterspeech but before they left the page, capturing their immediate reactions.
    \item Counterspeech robot username: To avoid anthropomorphisation and gendered connotations \cite{goversAIDrivenMediationStrategies2024}, we did not personify the moderation bot. Instead, we named it Civilbot (\begin{CJK*}{UTF8}{gbsn} 文明罗伯特 \end{CJK*} in Chinese). To further ensure transparency, we appended the tag platform to its username, clarifying that it was an official platform-moderation agent rather than a real user.
    \item Reply length: we adopted a soft-approximate approach by prompting GPT-5 to generate one short paragraph per speech act (detailed in Appendix \ref{app:prompt}). The resulting length variance allowed us to balance response lengths across different strategies.
\end{itemize}

\subsection{Participants}

\begin{table}[t]
    \centering
    \caption{Summary of Participant Demographics and Covariates ($N=52$)}
    \label{tab:demographics}
    \footnotesize
    \begin{tabularx}{\columnwidth}{@{} 
        l 
        >{\centering\arraybackslash}p{2.2cm} 
        X 
    @{}}
        \toprule
        \textbf{Covariates} & \textbf{Value Type} & \textbf{Values/Distribution} \\
        \midrule
        Gender & Categorical & \makecell[l]{Female ($n=27$) | Male ($n=25$)} \\
        \midrule
        Age & Continuous & \makecell[l]{Range (18--27)\\Mean = $20.17$, Std. = $2.52$} \\
        \midrule
        Education Level & Ordinal & \makecell[l]{High School ($n=1$)\\ Undergraduate ($n=40$)\\ Graduate ($n=11$)} \\
        \midrule
        Major & Nominal & \makecell[l]{Law ($n=11$) | Economics ($n=10$) \\ Engineering ($n=10$) | History ($n=7$) \\ Literature ($n=7$) | Science ($n=7$)} \\
        \midrule
        AI Literacy & Continuous \newline (Likert) & \makecell[l]{Scale: 1 (Very Low) to 5 (Very High) \\ Mean = $3.10$, Std. = $1.29$} \\
        \midrule
        AI Use & Continuous \newline (Likert) & \makecell[l]{Scale: 1 (Never) to 5 (Daily) \\ Mean = $4.37$, Std. = $0.87$} \\
        \bottomrule
    \end{tabularx}

    \smallskip
    \small
    Note: The last two covariates (AI Literacy and AI Use) are self-reported on a 5-point Likert scale.
\end{table}

Given that bystanders lack an accessible sampling frame and cannot be directly identified through platform-level data, we adopted convenience sampling \cite{golzarConvenienceSampling2022} by posting recruitment information in WeChat to reach individuals who were easy to access and willing to participate. We supplemented this with limited snowball sampling \cite{biernackiSnowballSamplingProblems1981}, which is commonly used when the target population is behaviour-defined and distributed within social networks. We recruited a total of 58 participants through WeChat, of which 52 were available for the quantitative analysis (demographics and individual differences shown in Table \ref{tab:demographics}). Prior to the formal experiment, all participants completed a pre-survey reporting their demographic information, social media usage habits, topic interests, and behavioural tendencies in response to hate incidents. We adopted a survey \cite{goversAIDrivenMediationStrategies2024} for participant inclusion and used self-report method \cite{jiaTacklingHateSpeech2025} to assess bystander roles. Specifically, the process was as follows: Participants' inclusion as active users of online communities was determined by collecting their social media usage frequency using a 5-point Likert scale. Subsequently, we assessed three key constructs, each measured by a separate 5-point Likert scale: interest in sensitive issues ("How interested are you in sensitive issues such as gender, race, region and so on?"), counter-engagement ("How often do you challenge or rebut hate speech?"), and hate speech endorsement ("How often do you like, share, or publish hate speech?"). Furthermore, in the formal experiment, bystander roles were reconfirmed through both the measurement of empathy in the pre-survey of each session and subsequent interviews. Based on these responses, we screened for our target group: potential bystanders, individuals who are active in online communities and interested in sensitive issues, but who typically remain silent and do not endorse hate speech. 

The required sample size was determined via a G*Power analysis for a repeated-measures ANOVA (within-subject design, 8 conditions). Based on the empirically derived median effect size \cite{sommetHowManyParticipants2023} ($Cohen's f = 0.175$), and a systematic review of reported effect sizes at CHI by \cite{ortloffSmallMediumLarge2025} (where the small–medium range of the type "human-centered computing" is approximately 0.10-0.26, making $0.175$ close to the median of this interval), we adopted $f = 0.175$ as the planned effect size for sample size estimation. This analysis indicated a minimum required sample size of 48 participants to achieve $80\%$ statistical power at $\alpha = .05$  \cite{cohenStatisticalPowerAnalysis1992}. Among the 58 participants we recruited, 5 completed a pilot study that informed revisions to the questionnaire, and 1 dataset was excluded due to incompleteness of all sessions, leaving 52 valid participants for analysis. The study protocol was approved by the university’s Institutional Review Board. Participants were informed  that they could withdraw at any time and would have access to psychological support if needed. The entire study lasted  approximately 45 minutes and consisted of four phases (see Fig. \ref{fig: experiment procedure}). Participants who completed all sessions received a  compensation of 50 RMB.

\subsection{Experiment procedure}

\begin{figure*}[t]
    \centering
    \makebox[\textwidth][c]{%
        \includegraphics[width=0.9\textwidth]{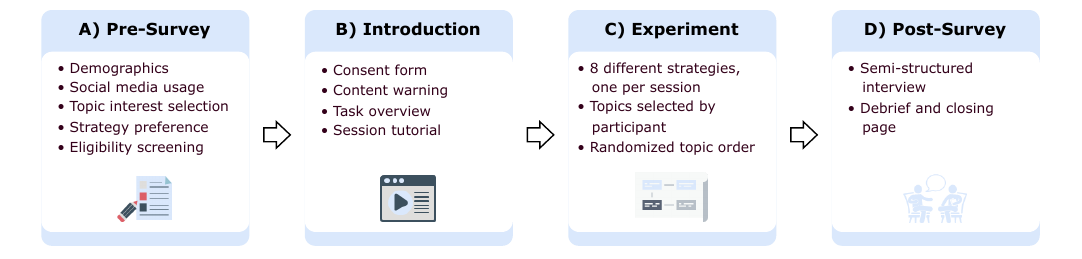}
    }
    \caption{The overall experiment procedure, including four phases: (A) Pre-survey, (B) Introduction, (C) Experiment sessions, (D) Post-survey.}
    \label{fig: experiment procedure}
\end{figure*}

Our study procedure contains four phases (see Fig. \ref{fig: experiment procedure}). In Phase A (Pre-survey), beyond the screening items, we included a self-developed strategy preference questionnaire to better understand participants and provide cues for subsequent experiments and interviews. This introduced three dimensions of counterspeech strategies and their bipolar expressions; participants indicated their preferences using a five-point scale. The main experiment comprised Phase B (Introduction) and Phase C (Experiment sessions). In Phase B, participants provided informed consent and read task instructions, which included a content warning for potentially offensive material. They then proceeded to Phase C, where participants freely selected eight questions of interest from the 27 items available. To mitigate potential topic effects and ensure comprehensive content coverage, the study incorporated three primary controls: The presentation order of the questions was randomized per participant, controlling for sequence effects. Second, participants freely chose their eight questions after browsing the full pool, ensuring broad topic engagement based on personal interest. Third, the strategy presentation order was also randomized, preventing the fixed coupling of any specific counterspeech strategy with a particular question. Each selected question corresponded to one session, and across the eight sessions, participants were exposed to all strategy types without repetition. Within each session, participants first encountered a piece of hate speech and completed the first survey measure. They were then presented with counterspeech generated by Civilbot, to which they could respond (e.g., by liking counterspeech or commenting on hate speakers). After clicking the "Home" button, the system displayed the second survey measure. In Phase D (Post-survey), participants engaged in semi-structured interviews. They were asked to reflect on their session experiences, discuss their subjective perceptions of Civilbot, elaborate on how Civilbot and these strategies affect their behaviours, and share their views on the role of Civilbot in shaping community social norms.

\subsection{Measurements}
During the initial scale design, we included multiple constructs to assess bystanders' evaluations before and after exposure to counterspeech, such as willingness to participate \cite{molinaRoleCivilityMetacommunication2018}, willingness to counterspeak \cite{sportelliLetsMakeDifference2025a}, counterspeech efficacy \cite{wachsEffectsPreventionProgram2023}, and empathy \cite{sportelliLetsMakeDifference2025a}. A pilot study (N=5) gathered qualitative feedback, which indicated item redundancy and the absence of a direct measure of counterspeech persuasiveness. These issues increased respondent fatigue and compromised measurement precision. We therefore streamlined items to reduce repetition and incorporated the well-validated Perceived Argument Strength Scale \cite{zhaoMeasurePerceivedArgument2011} to better capture argument quality. The final framework consists of three dimensions (detailed in Appendix \ref{app:questionnaire}): (1) perceived counterspeech quality (e.g., convincing and strong reasons), (2) subjective acceptance (e.g., credibility, importance, overall agreement), and (3) behavioural tendencies (e.g., confidence in countering, willingness to participate). Items for pre-post comparison (e.g., confidence in countering, willingness to participate) were measured at both time points to track changes, serving as a self-comparison baseline for assessing the intervention's effects. Items directly evaluating counterspeech were measured only post-exposure to avoid repetition. Empathy was assessed only in the pre-survey to exclude hate-endorsing respondents. Since the original scales were in English, we applied back-translation to ensure the conceptual validity of their Chinese version.

\subsection{Data Analysis}

To address the research questions, we employed a mixed-methods approach combining quantitative survey analysis with qualitative content analysis. For the quantitative component, we first calculated mean scores for each item to identify general trends in the effectiveness of all counterspeech strategies, providing a quantitative backdrop for subsequent qualitative analysis of RQ1. For RQ2, one-way ANOVA tests were conducted to examine the main effects of each of the three dimensions: intent, tone, and sentence type. Due to the conceptual distinctness of these dimensions, separate one-way ANOVAs were employed instead of MANOVA. Where significant main effects were identified, two-way ANOVA was performed to investigate potential interaction effects, followed by simple effects analysis where appropriate to uncover more granular patterns \cite{girdenANOVA1992}. Additionally, we conducted exploratory pairwise comparisons between all eight strategy groups via pairwise t-tests. These analyses were intended only to complement the ANOVA results by highlighting potential patterns that could inform our qualitative interpretation, rather than to establish definitive effects. Therefore, we did not apply corrections for multiple comparisons. The qualitative analysis of open-ended feedback followed the thematic analysis approach \cite{khokharTheoryDevelopmentThematic2020}, involving an iterative process wherein two researchers independently coded the data, identified recurring patterns, and through discussion consolidated them into salient themes reflecting participants' perceptions of Civilbot effectiveness, authenticity, contextual appropriateness, etc.

\section{Results}
\label{sec5 results}

\begin{figure*}[t]
    \centering
    \makebox[\textwidth][c]{%
        \includegraphics[width=0.85\textwidth]{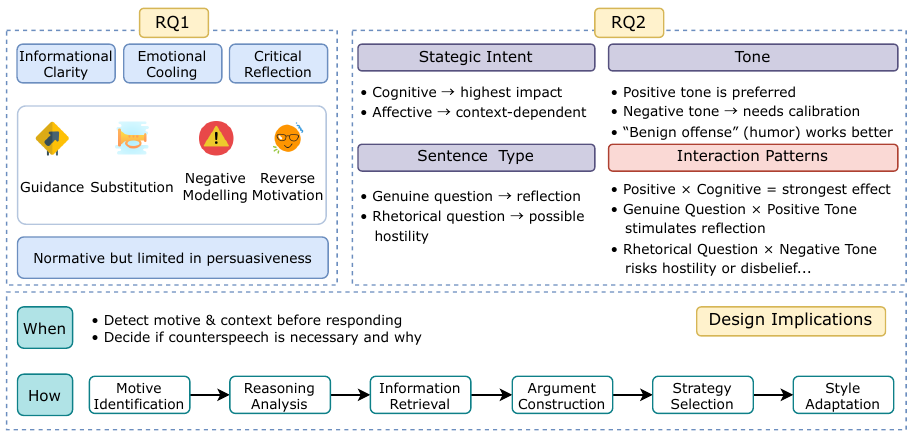}
    }
    \caption{Overview of results for RQ1–R3. RQ1 shows overall effects on bystanders, Civilbot's roles for them, and perceived community-level mechanisms; RQ2 presents strategy-level effects and key interaction patterns; Design implications illustrate design insights of Civilbot from the perspective of bystanders, organized by when to counterspeak and how to counterspeak.}
    \label{fig:results}
\end{figure*}

\subsection{RQ1: Counterspeech Chatbot's Overall Influence}

\begin{figure*}[t]
    \centering
    \makebox[\textwidth][c]{%
        \begin{subfigure}{0.42\textwidth}
            \centering
            \includegraphics[width=\linewidth]{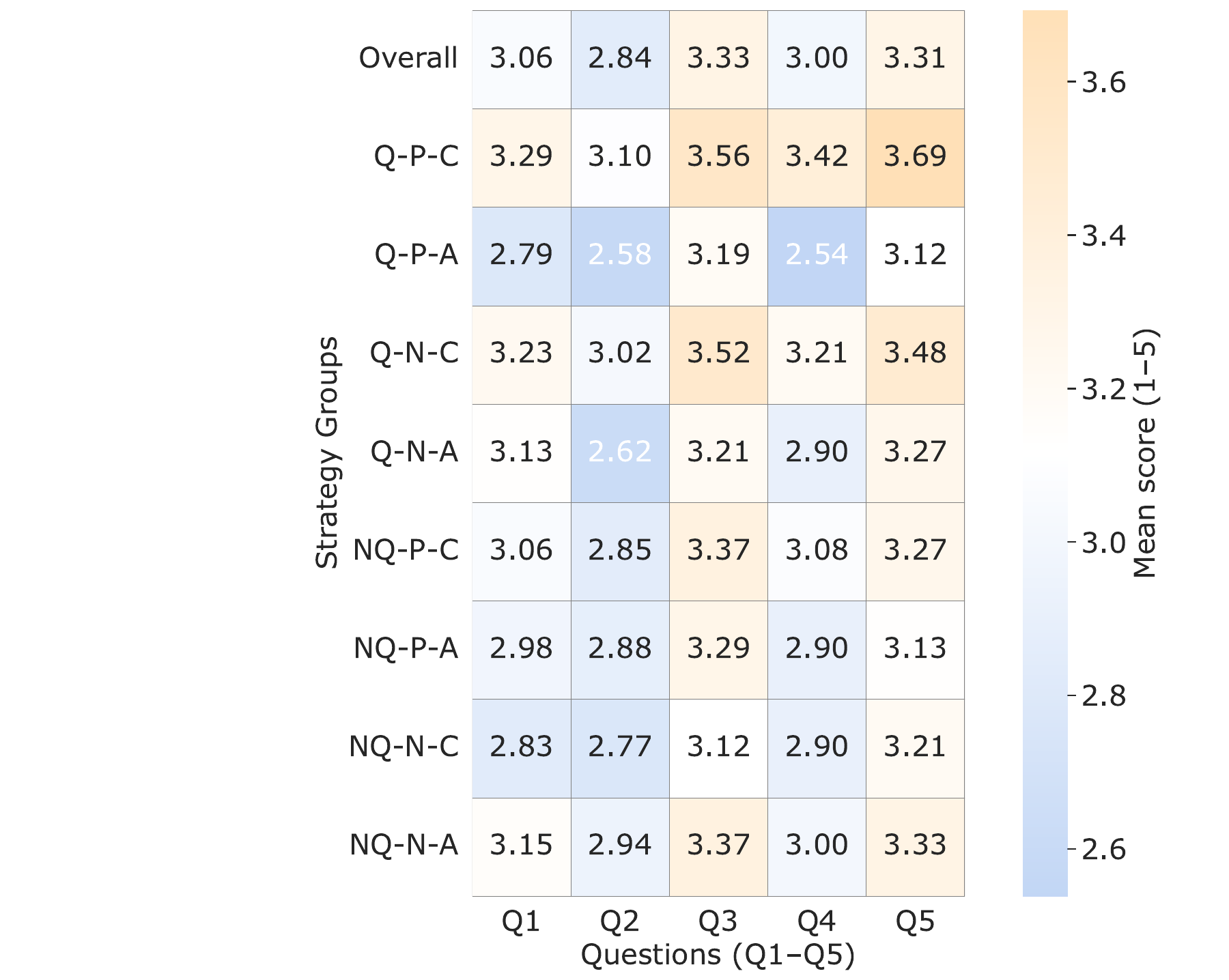}
            \caption{Mean Score of Q1-5}
            \label{fig:sub1}
        \end{subfigure}
        \hspace{0.04\textwidth}
        \begin{subfigure}{0.39\textwidth}
            \centering
            \includegraphics[width=0.55\linewidth]{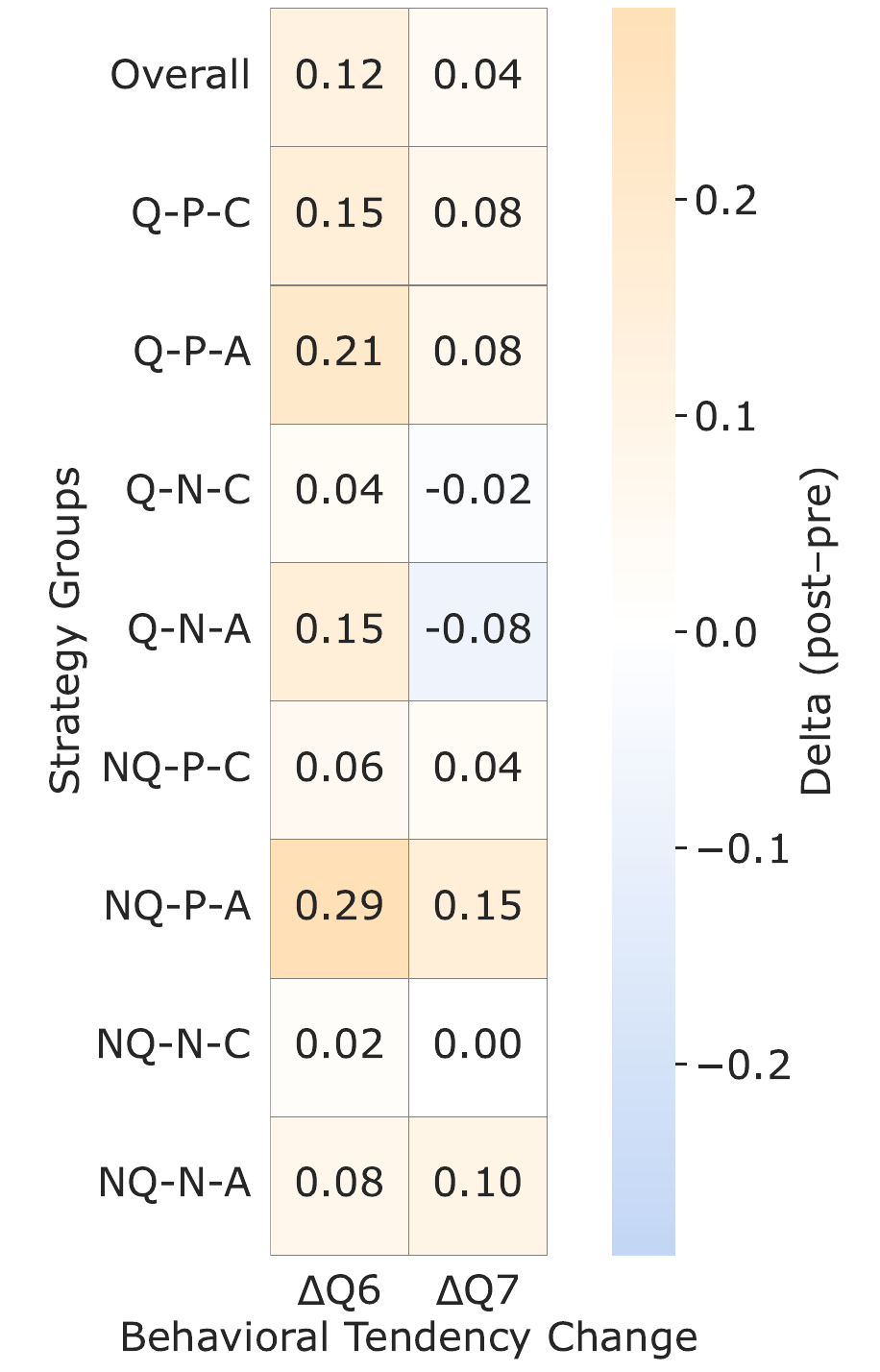}
            \caption{Mean Score of Delta Q6-7}
            \label{fig:sub2}
        \end{subfigure}
    }
    \caption{Heatmap of the correlation between mean scores of different variables and the eight strategy groups. On the y-axis, Q/NQ indicates question or non-question sentence type, P/N indicates positive or negative tone, and C/A indicates cognitive or affective strategic intent. Variables on the x-axis: Q1 (convincing reason), Q2 (strong reason), Q3 (credibility), Q4 (importance), Q5 (overall agreement), Q6 (confidence in countering) and Q7 (willingness to participant)}
    \label{fig:meanscore}
\end{figure*}

Overall, counterspeech delivered by Civilbot had a complex yet meaningful influence on bystanders. On the attitudinal level, it was generally perceived as credible and normative, reaffirming that hate speech is unacceptable and offering comfort to silent users. Yet its reasoning was often seen as shallow or "\textit{too AI-like}", limiting persuasiveness and its ability to mobilize engagement. On the behavioural level, its effects were subtle and sometimes contradictory—acting alternately as guidance, substitution, negative modelling, or reverse motivation, depending on users' self-efficacy and motivation. At the community level, Civilbot contributed to shaping the climate by informing bystanders, cooling their emotions, and encouraging them critical thinking. Together, these findings highlight both the potential and the limits of Civilbot in supporting bystander engagement and mediate online community climate.

\subsubsection{Perceived as Normative but Limited in Persuasiveness}
\label{sec5.1.1}
Bystanders basically perceived Civilbot as credible and normative but not strongly persuasive. Quantitatively (see Fig. \ref{fig:meanscore}), it received moderate ratings on \textit{credibility} (3.33/5) and \textit{overall agreement }(3.31/5), while scoring lower on \textit{convincing reason}, \textit{strong reason}, and \textit{importance}, with \textit{strong reason} in particular falling below the passing threshold (2.84/5).

Participants often described Civilbot as a transmitter of community norms (P20, P15, P5, P17, P48, P54) and even a representative of ethical values, especially for those with higher AI literacy. For some, its presence helped counterbalance despair in hostile comment sections: "\textit{If the comment section is full of irrational racist remarks, you might feel disillusioned about the world. But when AI [Civilbot] shows up, it reassures me that normal human values still exist}" (P21, P49, P54, P58). Others emphasized that because chatbots are perceived as relatively objective, their interventions carry stronger normative weight than those of ordinary users: "\textit{People usually won't argue with a robot. If it says something is problematic, most will accept it as correct}" (P30). Some participants even felt that siding with Civilbot placed them on moral high ground (P7, P12, P17). At the same time, views on its identity were divided: while some dismissed AI outputs as superficial recombinations (P1), others considered them a form of "\textit{collective intelligence}" (P6, P21), or simply cared more about content than source (P5, P27).

However, participants rated Civilbot low on perceived quality of counterspeech, largely because its arguments were perceived as formulaic, off-topic, or shallow (P1, P4, P5, P8, P11, P26, P27, P29, P31, P37, P45, P58). As one participant put it, "\textit{It didn't explain the causes of the issue or respond to the actual question—just gave a detached piece of advice}" (P26). Several emphasized that effective counterspeech requires engaging with the speaker's underlying logic rather than replying at the surface level. As P6 noted, "\textit{To refute someone, you first need to know their reasoning … it's about uncovering their motives and addressing their logic, rather than just throwing back a generic line.}" Such perspectives highlight that participants expected Civilbot to probe the motives behind hateful speech, exposing logical gaps or critical points that could make counterspeech more persuasive. Additionally, some criticized its "\textit{AI-like}" phrasing and formal tone (P8, P19, P41), which clashed with the informal style of hate speech, making responses feel distant or even unintentionally ridiculous (P5, P7, P13, P21, P22, P24, P26, P39, P40, P46). Several highlighted that counterspeech should adapt to the norms and dominant hate narratives of each platform (P7, P17, P18, P24, P34), and one even suggested adopting more implicit, moderate expressions aligned with Chinese cultural traditions (P10).

Interestingly, participants sometimes rate low due to perceived inappropriateness of rebutting certain posts. When hate speakers were merely sharing personal opinions with limited hostility, participants felt that counterspeech could appear excessive or misdirected, thereby creating unnecessary tension in an otherwise normal discussion space—"\textit{It really felt like attacking someone}" (P4). Others noted cases where Civilbot misinterpreted a post and delivered a counterspeech based on that misunderstanding (P21). Preferences also diverged: while some participants favoured addressing factual inaccuracies (P23), others argued that biased or emotionally charged comments likewise warranted a response (P11, P12).

\colorlet{framecolor}{boarder}
\colorlet{shadecolor}{box}
\setlength\FrameRule{0pt}
\begin{frshaded*}
\noindent In summary, participants generally viewed Civilbot as a legitimate normative voice, yet its persuasiveness was constrained by weak reasoning, rigid expression, and limited sensitivity to context. These findings highlight the need for future designs to balance normative authority with adaptive context-awareness.
\end{frshaded*}


\subsubsection{Mixed Behavioural Effects: Guidance, Substitution, Negative Modelling, and Reverse Motivation}
\label{sec5.1.2}
On the behavioural dimension, Civilbot showed only a modest positive effect on confidence in countering (Q6) and willingness to participate (Q7). This limited impact is understandable: behavioural change is shaped by complex factors such as knowledge, empathy toward targets, and communication habits, and is more likely to evolve over time. Interviews further revealed that Civilbot's influence on bystanders was subtle and often paradoxical, ranging from a guidance effect ("Civilbot did well, so I can join in"), to a substitution effect ("Civilbot did well, so I don't need to act"), a negative modeling effect ("Civilbot did poorly, so I lack confidence to respond"), and even a reverse motivation effect ("Civilbot did poorly, so I should step in to supplement"). These patterns are closely tied to the varied reasons bystanders choose silence in the first place.

As a \textbf{guidance}, Civilbot provided alternative perspectives that encouraged reflection and sometimes lowered the threshold for participation. Participants who felt under-informed or cautious about debating hate speech noted that chatbot interventions helped keep the conversation alive: "\textit{Its comments can serve as a starting point, attracting more people to join in and keeping an anti-hate atmosphere}" (P10). Others felt inspired by specific arguments, which provided cognitive scaffolding for their own contributions (P40, P55). For instance, in a scenario involving hate speech accusing Koreans of being stingy and arrogant, P27 noted: "\textit{It reminded me not to generalize about all Koreans, and then I thought of examples to use in my own counterspeech.}" Similarly, Civilbot’s responses sparked empathy toward targeted groups (P4) or curiosity about hate speakers’ motives (P6, P52), thereby motivating engagement. In a case attacking nurses for alleged poor conduct, P4 reflected: "\textit{At first I thought nurses had nothing to do with me, but after Civilbot’s response, I realized how unfair it was, especially remembering their effort during COVID-19}". In these instances, Civilbot’s intervention served to remind participants of their social accountability or offered heuristics on how to construct an effective counterspeech.

Civilbot also acted as a \textbf{substitution}. For participants who opposed hate speech but feared conflict, its presence reduced the pressure to respond personally. Many described "\textit{liking counterspeech comments}" as a low-cost way to register opposition without direct confrontation (P4, P7, P11, P16, P17, P18, P24, P32, P48). "\textit{I don't want to be the pioneer or opinion leader. I just want my like to show the hate speaker they're wrong, and I hope the counterspeech comments get tens of thousands of likes while the hate speech gets only a few}" (P4). Others even wished Civilbot could serve as their spokesperson, absorbing personal viewpoints and expressing them on their behalf (P10, P57). Participants highlighted that, unlike humans, chatbots never tire, remain emotionally unaffected, and can consistently respond to recurring hate topics—an advantage for long-term engagement (P2, P5, P18, P36). For many, Civilbot's presence provided reassurance that they were part of a larger anti-hate majority (P20, P33, P54, P56, P57).

At the same time, Civilbot sometimes functioned as a \textbf{negative modelling} or \textbf{reverse motivation}. When its interventions were ineffective, some participants perceived it as a negative modelling and reported decreased confidence in their own ability to intervene. For example, regarding hate speech labeling Asian Americans as "weak", P23 noted: "\textit{AI knows this topic better than me, yet it performs poorly… I don’t know what to do.}" Yet others felt compelled to "step in" precisely because Civilbot's contribution fell short: "\textit{After reading its comment, I wanted to add my own—not because it was good, but because it missed the point, and I couldn't resist correcting it}" (P31). In this way, Civilbot inadvertently created a kind of psychological safety net, acting as a first responder that reduced the perceived risk of joining in. This divergence likely stems from the interplay between the participant's self-efficacy in countering and their perceived efficacy of Civilbot. When Civilbot performs poorly, participants with higher self-confidence—often linked to topic familiarity, writing capability, or the perceived weakness of the hate speech—feel motivated to correct the error. In contrast, those with lower confidence may interpret Civilbot's failure as a signal of the task's difficulty, leading to withdrawal.

Finally, participants noted that the behavioural impact of counterspeech may be constrained by external factors such as content moderation. Some worried that over-policing could mistakenly punish counterspeech (P9), underscoring the need for clearer distinctions between hate speech and counterspeech. At the same time, they emphasized that Civilbot's role is not to compel or entice every bystander to actively respond. It establishes a reliable baseline of opposition to hate, providing essential reassurance and symbolic justice even for bystanders who prefer indirect forms of participation, such as liking a counterspeech comment or simply observing its presence (P24).

\colorlet{framecolor}{boarder}
\colorlet{shadecolor}{box}
\setlength\FrameRule{0pt}
\begin{frshaded*}
\noindent In summary, Civilbot's behavioural influence is complex and nuanced. It can inspire reflection and participation, substitute for silent bystanders, undermine confidence through weak responses, or motivate corrective action when it fails. Its value lies not in mobilizing all audiences to counterspeak, but in sustaining a minimal yet consistent countervoice that anchors community norms and offers bystanders a safer space to position themselves.
\end{frshaded*}

\subsubsection{Shaping Community Climate through Informational Clarity, Emotional Cooling, and Critical Reflection}
\label{sec5.1.3}
Beyond individual actions, participants emphasized Civilbot's broader role in shaping community climate. They valued its ability to provide timely information and perspectives, preventing hateful misinformation from misleading uninformed bystanders, especially vulnerable groups such as teenagers (P9, P15, P20, P26, P32). As one participant noted, "\textit{Its greatest value is helping neutral people realize the truth}" (P32). Because the information was presented as reasoning rather than instructions, some participants felt it was more credible—they could follow the logic themselves and thus trust their own conclusion (P23), and several even reported learning laws, theories, or statistics from its comments (P13, P20, P21, P30, P56).

Civilbot also helped stabilize emotions and discourse. Counterspeech could cool overheated reactions, interrupt escalating disputes, and prevent large-scale mobilization of hate, partly because hate speakers were unlikely to engage directly with a chatbot (P24, P29, P52). For some (P20, P56), this provided emotional relief: "\textit{At first I believed the hateful comment and felt angry at Koreans, but Civilbot's response calmed me down and reminded me of a more positive perspective}".

Finally, Civilbot fostered critical reflection and dialogue by presenting opposing viewpoints, which encouraged bystanders to contribute. As one participant observed, "\textit{When two opposite opinions are put forward, people naturally start to discuss around them}" (P19). In this way, counterspeech functioned not only as correction but also as a catalyst for community-level discussion (P22, P28, P43, P47, P56).

\colorlet{framecolor}{boarder}
\colorlet{shadecolor}{box}
\setlength\FrameRule{0pt}
\begin{frshaded*}
\noindent In summary, Civilbot shaped the community climate by (1) offering information that supports recognition of hate and misleading cognition, (2) cooling emotions and preventing escalation, and (3) stimulating dialogue through diverse perspectives. These functions positioned it less as a debater and more as a balancer of knowledge, emotion, and reflection, thereby helping consolidate community values at a broader level.
\end{frshaded*}

\subsection{RQ2: Influences between Counterspeech Strategies}
\label{sec5.2}

\begin{table*}[t]
  \caption{Main Effects of the Three Counterspeech Dimensions across the Three Measurements}
  \label{tab:main_effects}
  \centering

  \makebox[\textwidth][c]{%
    \begin{tabular}{
      l l
      S[table-format=2.3]
      S[table-format=1.3]
      S[table-format=1.3]
    }
      \toprule
      {Measurement Direction} & {Factor (Counterspeech Dimension)} &
      {$F$-value} & {$p$-value} & {$Cohen's f$} \\
      \midrule
      \multirow{3}{*}{Perceived Quality}
        & Sentence & 9.096 & \textbf{0.003**} & 0.148 \\
        & Tone     & 2.659 & 0.104 & 0.080 \\
        & Intent   & 18.589 & \textbf{$<$0.001***} & 0.212 \\
      \midrule
      \multirow{3}{*}{Subjective Acceptance}
        & Sentence & 5.189 & \textbf{0.023*} & 0.112 \\
        & Tone     & 0.313 & 0.576 & 0.027 \\
        & Intent   & 24.474 & \textbf{$<$0.001***} & 0.243 \\
      \midrule
      \multirow{3}{*}{Behavioural Tendency}
        & Sentence & 0.011 & 0.915 & 0.005 \\
        & Tone     & 1.662 & 0.198 & 0.063 \\
        & Intent   & 0.564 & 0.453 & 0.037 \\
      \bottomrule
    \end{tabular}
  }

  \smallskip
  \small
  Note: * $p < .05$, ** $p < .01$, *** $p < .001$.
\end{table*}

\begin{table*}[t]
  \caption{Interaction Effects between Counterspeech Dimensions across the Three Measurements}
  \label{tab:interaction_effects}
  \centering

  \makebox[\textwidth][c]{%
    \begin{tabular}{
      l l
      S[table-format=1.3]
      S[table-format=1.3]
      S[table-format=1.3]
    }
      \toprule
      {Measurement Direction} & {Interaction (Counterspeech Dimension combination)} &
      {$F$-value} & {$p$-value} & {$Cohen's f$} \\
      \midrule
      \multirow{3}{*}{Perceived Quality}
        & Sentence $\times$ Intent & 0.022 & 0.882 & 0.007 \\
        & Sentence $\times$ Tone   & 0.472 & 0.492 & 0.034 \\
        & Intent $\times$ Tone     & 4.756 & \textbf{0.030*} & 0.107 \\
      \midrule
      \multirow{3}{*}{Subjective Acceptance}
        & Sentence $\times$ Intent & 0.188 & 0.665 & 0.021 \\
        & Sentence $\times$ Tone   & 0.860 & 0.354 & 0.046 \\
        & Intent $\times$ Tone     & 1.071 & 0.301 & 0.051 \\
      \bottomrule
    \end{tabular}
  }

  \smallskip
  \small
  Note: * $p < .05$, ** $p < .01$, *** $p < .001$.
\end{table*}

\begin{table*}[t]
  \caption{Simple Effects Analysis of Intent at Different Levels of Tone (Perceived Quality)}
  \label{tab:simple_effects}
  \centering

  \makebox[\textwidth][c]{%
    \begin{tabular}{
      l
      S[table-format=2.3]
      S[table-format=1.6]
      S[table-format=1.6]
    }
      \toprule
      {Comparison} & {$F$-value} & {$p$-value} & {$Cohen's f$} \\
      \midrule
      Positive-Affective vs Positive-Cognitive & 24.198 & \textbf{$<$0.001***} & 0.343 \\
      Negative-Affective vs Negative-Cognitive & 2.079 & 0.151 & 0.100 \\
      \bottomrule
    \end{tabular}
  }

  \smallskip
  \small
  Note: * $p < .05$, ** $p < .01$, *** $p < .001$.
\end{table*}

RQ2 revealed differentiated effects of counterspeech strategies across perceived quality, subjective acceptance, and behavioural tendencies. Among the three measurements, intent proved most decisive: cognitive strategies generally outperformed affective ones, though the latter retained situational value, especially when sequenced adaptively—for instance, a positive affective move could ease emotional tension before a cognitive argument. Tone shaped behaviour more than perception. Positive tone enhanced confidence in countering and willingness to participate, while negative tone required careful calibration, working best as "benign offense" (e.g., humor) rather than direct attack. Question forms showed significant effects overall but carried nuanced potential: genuine questions could stimulate reflection, whereas rhetorical ones risked hostility or disbelief. Importantly, interaction effects revealed that positive tone amplifies the strength of cognitive strategies, while affective strategies often falter in this register.


\subsubsection{Strategic Intent Shapes Perceived Quality, Subjective Acceptance, and Behavioural Tendencies}

Participants in interviews consistently valued cognitive strategies that encouraged reasoning through offering facts and correcting misconceptions. These strategies aligned with their expectation that Civilbot should provide knowledge to prevent the spread of harmful narratives and to stimulate reasoned debate. This finding was also consistent with the quantitative results, which indicated that strategic intent significantly influenced perceived quality ($F=18.59, p<0.001$, $\text{Cohen's } f=0.21$) and subjective acceptance ($F=24.47, p<0.001$, $\text{Cohen's } f=0.24$) (see Table \ref{tab:main_effects}). Suggestions for improvement included digging deeper into the motives of hate (P7, P21, P23), using logical and focused arguments (P1, P6, P11, P12, P17, P29, P30, P34, P37, P52), combining theory with concrete cases (P5, P9, P17, P19, P22, P24, P27, P30, P32, P33, P36), and even incorporating external links (P22, P23, P28, P31) or visual aids (P9). As one participant put it, "\textit{Charts would catch more attention}" (P9).

By contrast, affective strategies elicited more nuanced and sometimes polarized reactions. For some, personal preference dictated rejection of emotional appeals, as they favored objective debate (P4, P9). However, these participants occasionally rated affective counterspeech highly when it successfully evoked empathy through meaningful associations—such as recalling the sacrifices of nurses during COVID-19 (P4), connecting to media portrayals (P9), or invoking shared human values like mutual respect (P5, P40, P53). These examples suggest that affective strategies can work when they trigger authentic connections rather than relying on generic moralizing. Other participants, however, found affective strategies expressed by a digital chatbot less acceptable, perceiving them as hollow or even absurd (P42, P45). For example, statements like "I stand with X" were dismissed as meaningless because Civilbot "\textit{is just a void machine}" without the capacity for real-world solidarity: "\textit{I don't care who it stands with; it's just typing words. If a real person says it, I believe they mean it. But the robot? It changes nothing}" (P9). Such responses highlight that emotional expressions risk backfiring by unintentionally reminding participants of Civilbot’s non-human identity, which may ultimately undermine credibility instead of fostering closeness. Similar scepticism extended to first-person pronouns, which some saw as dissonant when used by an AI-driven chatbot (P7).

\colorlet{framecolor}{boarder}
\colorlet{shadecolor}{box}
\setlength\FrameRule{0pt}
\begin{frshaded*}
\noindent Taken together, participants suggested that cognitive and affective strategies are not mutually exclusive but rather complementary. Cognitive approaches are necessary when clear misconceptions must be corrected (P18, P20), but affective appeals may be effective in engaging hostile speakers or indifferent bystanders (P5). When paired with a negative tone, affective strategies were considered more persuasive if delivered after factual reasoning, so as not to appear as mere venting (P26). Ultimately, participants envisioned an adaptive combination of strategies, tailored to the audience and context, as the most effective form of counterspeech (P18, P20).
\end{frshaded*}

\subsubsection{Tone Influences Behavioural Tendencies, but Its Effect Depends on Context}
\label{sec5.2.2}

Overall, participants tended to favour a positive tone, despite conceding that a negative tone was more effective at capturing attention. Conditions such as Q-P-A and NQ-P-A scored relatively high in behavioural items (see Fig. \ref{fig:meanscore}). This pattern, as revealed by the qualitative data, suggests potential reasons. First, if Civilbot counters hate with hostility, it risks normalizing aggression and even raising ethical concerns about "\textit{a machine attacking a human}", especially when the hateful comment is not strongly malicious (P9, P11, P21, P22, P26, P29, P55). Second, negative tones may contribute to hate speakers' resistance, escalate conflicts, and inadvertently harm innocent bystanders or cautious victims who are already vulnerable in public discussions (P6, P11, P14, P16, P20, P22, P23, P30, P32, P43). However, positive tones are not a universal solution: when paired with extreme hate, they risk seeming absurdly mismatched and failing to engage anyone (P9, P28, P38). 

Some participants emphasized that negative tones, while potentially effective, should be employed with restraint and caution. A degree of sharpness could help capture attention, convey emotion such as a sense of justice, and assert community norms. As P13 explained: "\textit{A stronger tone like this can directly bring more people who are watching into an emotion like yours, and it can make them choose to stand on the side with a stronger tone towards you. They may be more inclined to believe it.}" Yet when overused, negativity risked souring the atmosphere and discouraging participation in discussions (P7, P10, P21, P25, P56, P57).

\colorlet{framecolor}{boarder}
\colorlet{shadecolor}{box}
\setlength\FrameRule{0pt}
\begin{frshaded*}
\noindent Ultimately, participants called for context-sensitive tone management. When hate is relatively weak, a gentle tone can reassure bystanders that meaningful dialogue is possible (P5). A balanced mix of positivity and negativity was considered ideal, drawing attention without succumbing to toxic dynamics. Some highlighted "benign offense", such as humorous sarcasm, as a particularly effective middle ground: "\textit{Although I [Civilbot] am attacking you, the ‘attack' is in quotes—it's playful, so you can attack me back. That makes it a closer form of exchange}" (P17).
\end{frshaded*}

\subsubsection{Question Forms May Encourage Reflection but Risk Backfiring}

Question-based sentence types were identified by participants as a potentially effective counterspeech strategy, and specific observations were made regarding the further influence of the question's precise form and matched tone. This finding was also consistent with the quantitative results, showing that sentence type significantly affected two measurements: perceived quality ($F=9.10$, $p=0.002$, $\text{Cohen's} f=0.15$) and subjective acceptance ($F=5.19$, $p=0.023$, $\text{Cohen's} f=0.11$). Genuine questions were valued for inviting reflection and broadening dialogue, often pairing well with positive tones to create a persuasive, approachable style (P7, P26, P42, P53). By presenting multiple possibilities and adding new information, such questions could encourage critical thinking among bystanders. In contrast, rhetorical or confrontational questions functioned less as invitations to reasoning and more as attacks. These were frequently perceived as rude or untrustworthy, yet when combined with negative tones, they conveyed emotional intensity and appeared forceful (P9, P20, P47).

Participants also noted that questions could help uncover the logic or motives behind hate speech, exposing inconsistencies (P15, P22, P28, P46). However, others warned that probing too deeply might backfire: hate speakers could use the opportunity to cite further evidence, reinforcing stereotypes if counterspeech failed to respond effectively (P22, P56). This illustrates the double-edged nature of questioning. While critical inquiry carries risks, participants stressed that its value lies less in "defeating" hate speakers than in sustaining constructive discussion. From this perspective, withholding responses for fear of giving opponents more space is counterproductive, as counterspeech can still guide bystanders and prevent dialogue from being dominated by hate. Therefore, Civilbot is expected not just to confront, but to supplement, guide, and mediate.

\colorlet{framecolor}{boarder}
\colorlet{shadecolor}{box}
\setlength\FrameRule{0pt}
\begin{frshaded*}
\noindent In short, quantitative results showed significant effects, and qualitative analysis revealed that questions can either stimulate critical reflection or undermine credibility, depending on how they are designed. For Civilbot, questions work best as tools for providing information, guiding reflection, and sustaining dialogue, rather than as blunt instruments of confrontation.
\end{frshaded*}

\subsubsection{Positive Tone Amplifies the Advantage of Cognitive Strategies}

Participants pointed to a potential relationship between tone and cognitive strategies. Cognitive strategies delivered in a negative tone risked being misinterpreted as emotional venting. While a negative tone could attract attention (P7, P9, P10, P21, P57), this attention might not be directed towards the reasoning itself, and might even be counterproductive. As P26 noted: "\textit{You need to be mindful of your [Civilbot's] tone. If you come in strongly with resentment, others may not have the patience to read what you specifically said, believing it's just venting.}" This perceived departure from neutrality, akin to "\textit{personal expression}", might further undermine Civilbot's credibility as a symbol of community social norms (P16, P22). Conversely, the combination of a positive tone with cognitive strategies raised the hope of constructive discourse (P5): it clarifies or provided information as material for discussion while the positive attitude signalled the potential for further communication rather than confrontation. As P33 pointed out: "\textit{Reasoning need to be delivered with a rational attitude.}" These findings are consistent with the Interaction analyses (see Table \ref{tab:interaction_effects}), which revealed that in perceived quality, a significant interaction was observed ($F=4.76, p=0.030, \text{Cohen's } f=0.11$). Furthermore, subsequent Simple-effect tests echoed this pattern: under positive tone, cognitive strategies substantially outperformed affective ones ($F=24.20, p<0.001, \text{Cohen's } f=0.34$), whereas under negative tone the two did not differ ($F=2.08, p=0.150, \text{Cohen's } f=0.10$). Participants explained that positive-tone cognitive counterspeech was perceived as more objective, fair, and conducive to discussion (P12, P16, P22). In contrast, affective strategies often struggled to evoke genuine empathy in a positive register (as discussed in Section \ref{sec5.2.2}), which might weaken their persuasive power.

\colorlet{framecolor}{boarder}
\colorlet{shadecolor}{box}
\setlength\FrameRule{0pt}
\begin{frshaded*}
\noindent Overall, these findings indicate that cognitive strategies are amplified by positive tone, which reinforces their credibility and constructive potential. Affective strategies, by contrast, do not enjoy the same boost and may come across as superficial when expressed positively. Under negative tone, strong emotional expression tends to overshadow strategic intent, blurring distinctions between cognitive and affective strategies. For Civilbot, this underscores the importance of aligning intent with tone: negative tone must be used with caution—while it can sometimes blunt strategic differences and undermine persuasion, it may also, in certain contexts, intensify affective influence, whereas positive tone should be prioritized to maximize the impact of cognitive counterspeech.
\end{frshaded*}

\subsubsection{Exploratory Analyses: Positive-Affective Questions, Non-Questions, and Human-AI Dynamics}

\begin{figure*}[t]
    \centering
    \makebox[\textwidth][c]{%
        \begin{minipage}{0.92\textwidth}
            \centering
            \begin{subfigure}{0.45\textwidth}
                \centering
                \includegraphics[width=\linewidth]{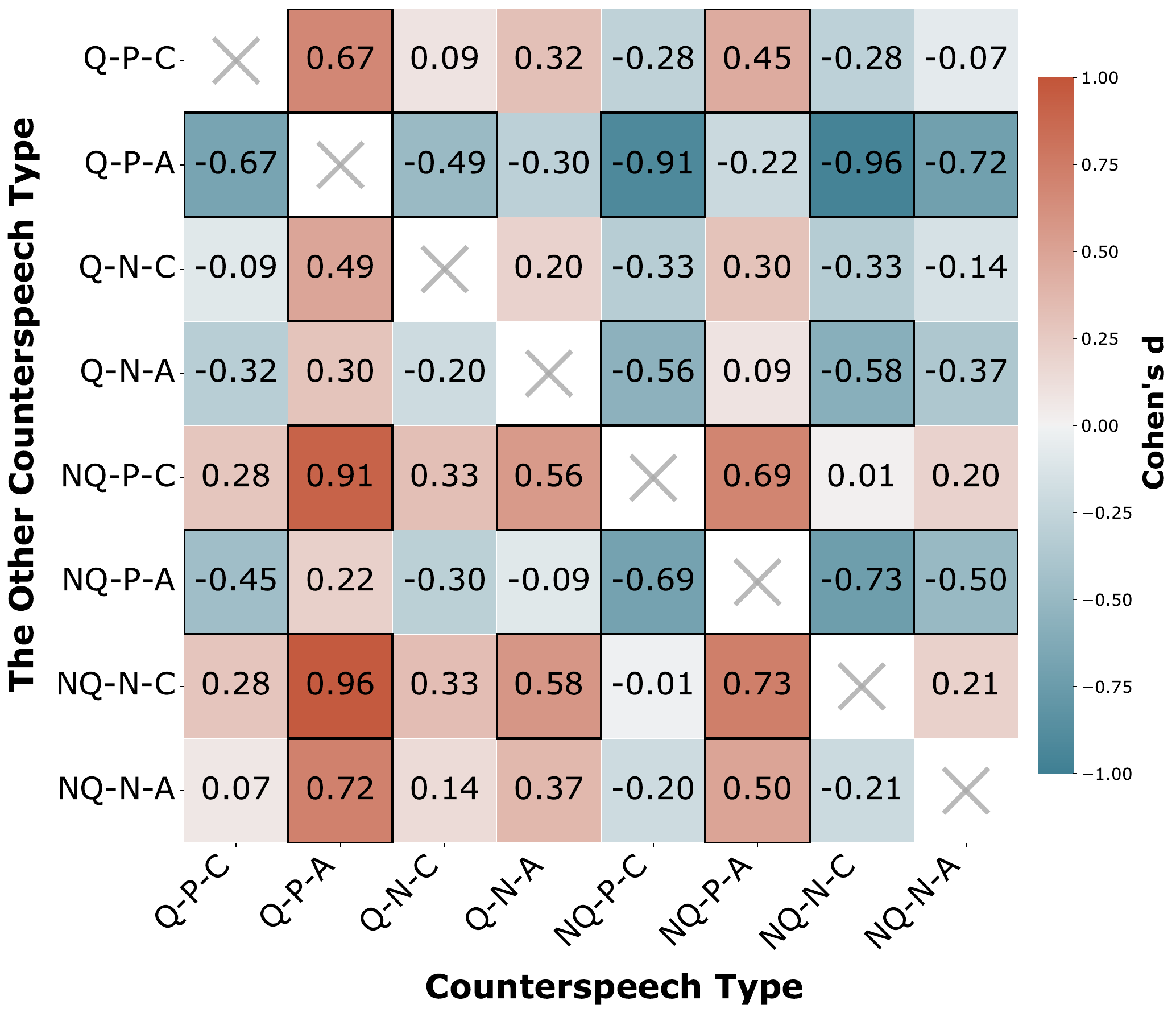}
                \caption{Pairwise Comparison in Perceived Quality}
                \label{fig:sub1}
            \end{subfigure}
            \hfill
            \begin{subfigure}{0.45\textwidth}
                \centering
                \includegraphics[width=\linewidth]{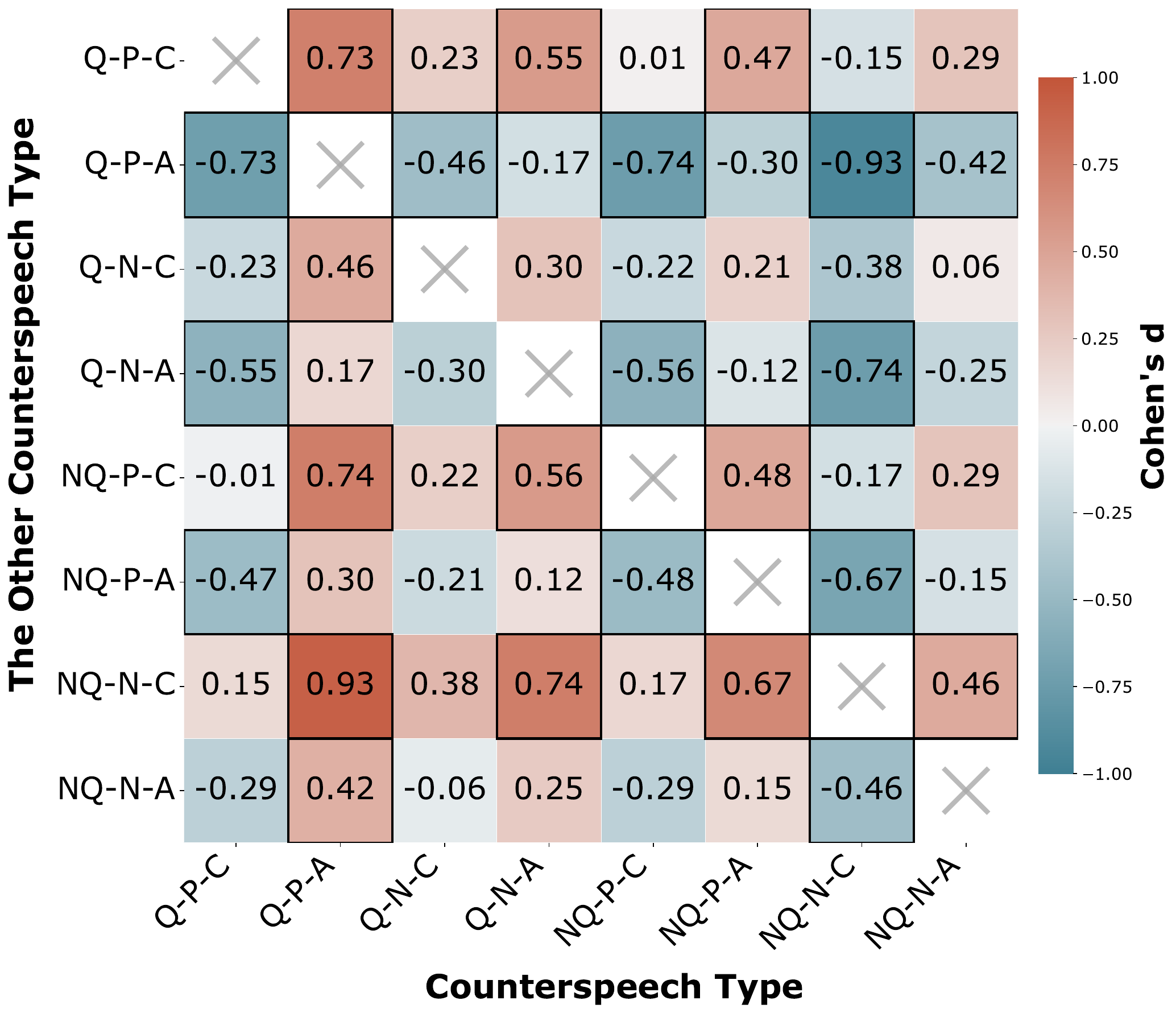}
                \caption{Pairwise Comparison in Subjective Acceptance}
                \label{fig:sub2}
            \end{subfigure}

            \vspace{1em}

            \begin{subfigure}{0.45\textwidth}
                \centering
                \includegraphics[width=\linewidth]{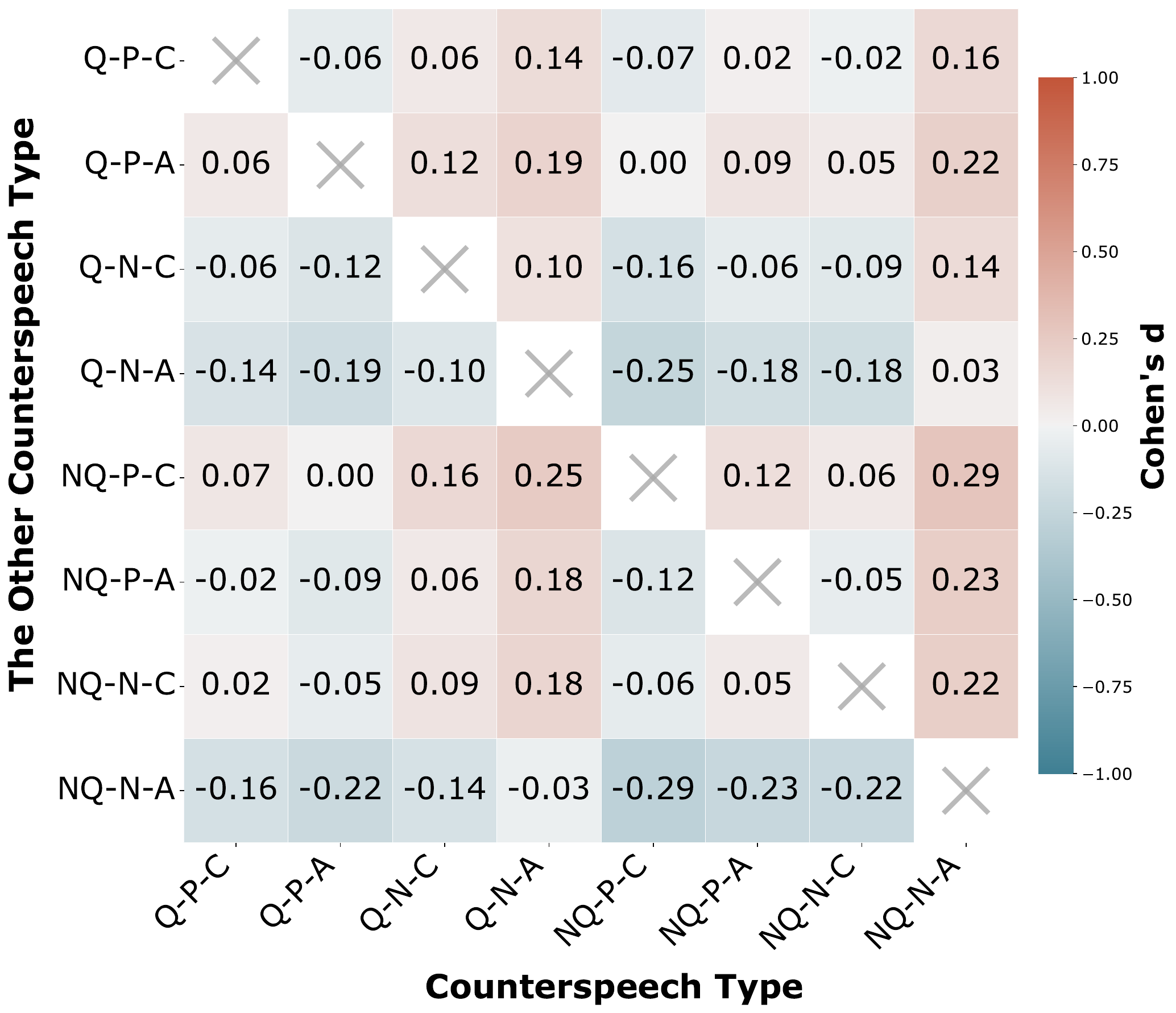}
                \caption{Pairwise Comparison in Behavioural Tendencies}
                \label{fig:sub3}
            \end{subfigure}
        \end{minipage}
    }
    \caption{Heatmap of pairwise paired t-tests between counterspeech types across the three questionnaire measures. Significant comparisons are outlined with bold borders. Cell values represent effect sizes (Cohen’s d). Perceived quality = mean(Q1–2); subjective acceptance = mean(Q3–5); behavioral tendencies = mean($\Delta$Q6–7) ($\Delta$ = post $-$ pre).}
    \label{fig:pairwise}
\end{figure*}

\begin{figure}[h!] 
    \centering 
    \begin{subfigure}{0.45\textwidth} 
        \centering
        \includegraphics[width=\linewidth]{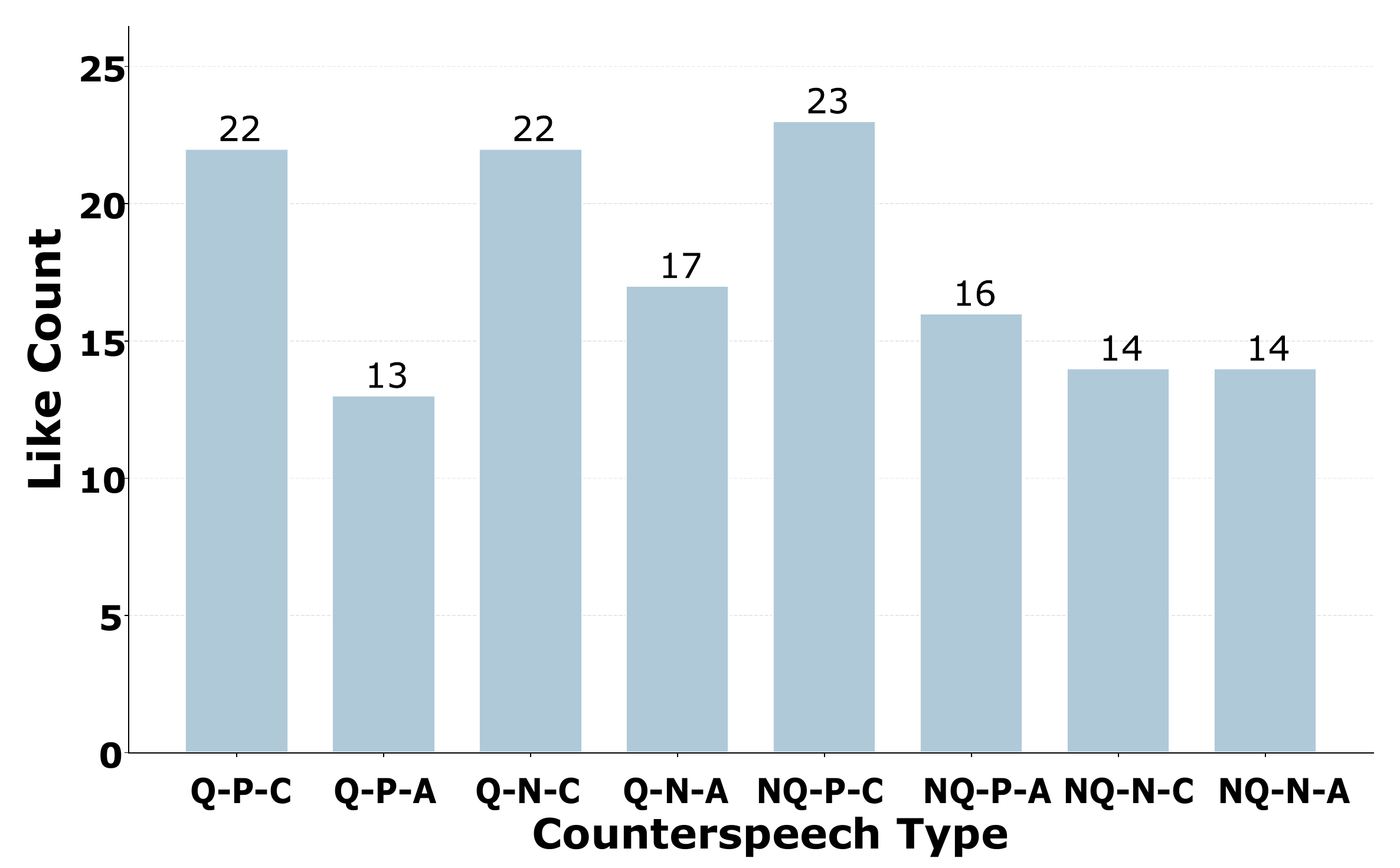} 
        \caption{Distribution of Likes} 
        \label{fig:sub1}
    \end{subfigure}
    \hfill 
    \begin{subfigure}{0.45\textwidth}
        \centering
        \includegraphics[width=\linewidth]{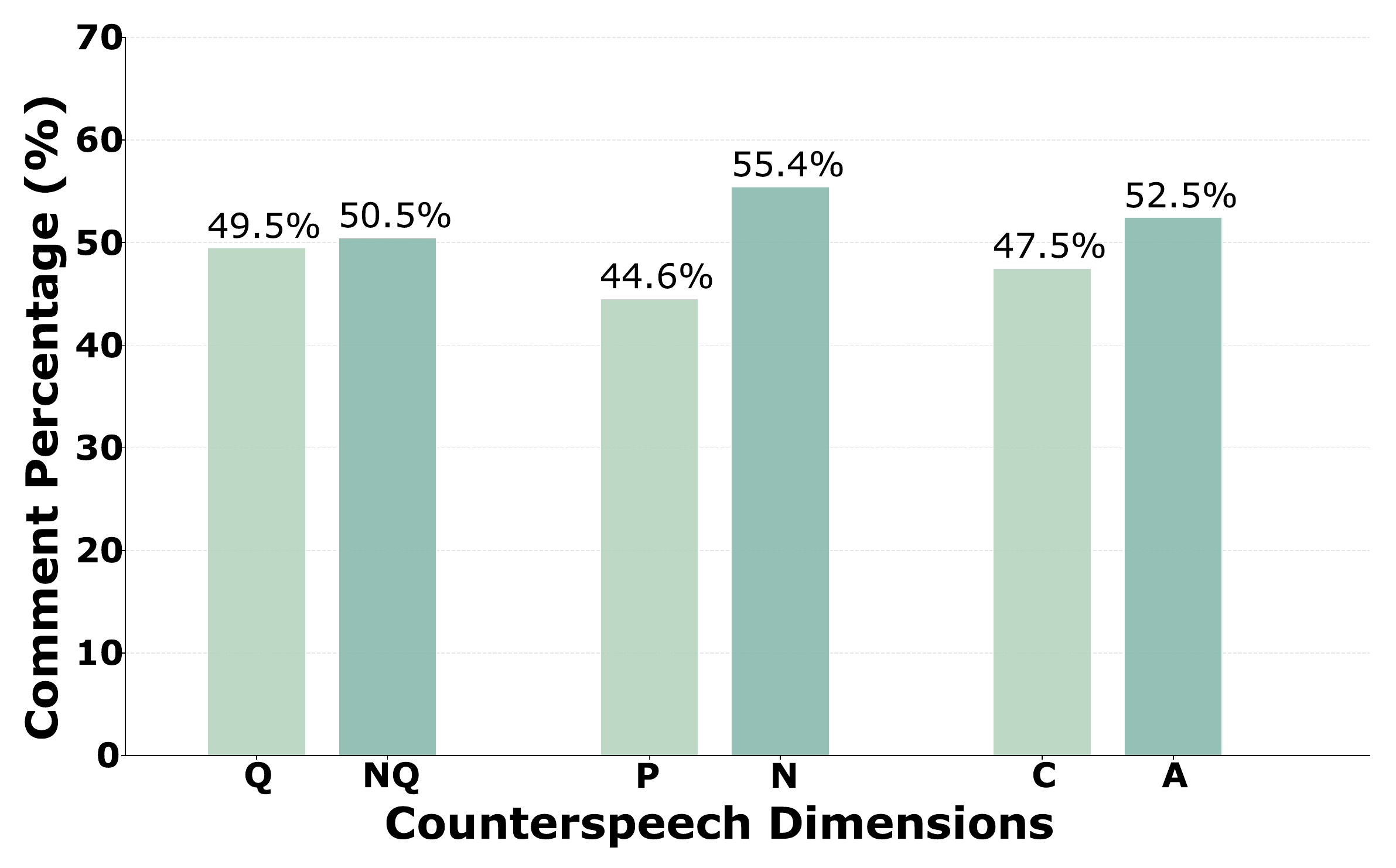}
        \caption{Distribution of Comments} 
        \label{fig:sub2}
    \end{subfigure}
    \caption{Distribution of participants' interactions across different counterspeech strategies.} 
    \label{fig:comments&likes} 
\end{figure}

Pairwise t-tests across the eight strategy groups revealed several exploratory patterns (see Fig.\ref{fig:pairwise}), which should be read as suggestive rather than causal.

In perceived quality and subjective acceptance, Q-P-A (positive-tone affective questions) performed worse than most other groups, with lower ratings of quality, credibility, importance, and agreement. Participants often perceived such utterances as idealized appeals that lacked authentic emotional resonance. On the one hand, positive-tone affective strategies were seen as too mild for condemnation yet insufficient for empathy. On the other hand, positive-tone questions leaned toward open-ended inquiry without adding substantive reasoning. As a result, this combination was generally weak, though some still valued its normative stance when responding to mild hate speech: "\textit{at least it conveys the right idea}" (P9, P20, P21, P27). Overall, NQ-N-A (non-question, negative, affective) performed relatively worst, yet relatively positive on willingness to participate (Fig. \ref{fig:meanscore}). This apparent contradiction aligns with earlier findings in Section \ref{sec5.1.2} that poorly received counterspeech can serve as negative exemplars, motivating participants to craft their own counterspeech.

Additional analyses of likes and participant-generated counterspeeche further nuanced this picture (see Fig. \ref{fig:comments&likes}). Q-P-A again received the fewest likes, consistent with its weak reception. Interestingly, participants' own counterspeech showed a preference of negative tone, diverging from their expectations of Civilbot. This reflects Civilbot's dual role: as a non-human chatbot and a symbolic actor of social norms, it is expected to remain objective and norm-affirming, avoiding slightly harsh negative tone. However, it is contrasted with human counterspeech, where negative tone is often valued as a tool of justice and emotional release. For participants, harsh counterspeech serve pragmatic purposes—raising the cost of hate (e.g., by making it risky or unpleasant to speak), disrupting hate speakers' goals of silencing targets, and in some cases even shifting or suppressing hate speakers (P5, P7, P16, P17, P19, P20, P24, P26). As one participant explained, "\textit{my own harsher counterspeech helps me vent, but I'd want AI to play a supporting role}" (P24).

\section{Discussion}
\label{sec6 discussion}
In this section, we integrate the findings and derive design implications for Civilbot. Specifically, we discuss two core dimensions: when to intervene—that is, whether counterspeech should be deployed and for what reasons; and how to intervene—that is, how counterspeech reasoning, evidence, and organization should be constructed. We also discuss the role of identity in the influence of chatbot counterspeech, and conclude with limitations and future directions.

\subsection{Boundaries of Chatbot Counterspeech: When and Why to Intervene}
\label{sec6.1 when}
Our results highlight participants' views on the scope of counterspeech and its complementarity with content moderation. In cases of extreme hate speech, counterspeech was often perceived as ineffective or even harmful. This is because posts containing slurs or large-scale harassment often leave little room for discussion; instead, counterspeech may risk reinforcing their salience, distract attention from urgent responses \cite{cepollaroCounterspeech2023}, and even accelerate their spread due to platform recommendation algorithms. As P4 noted, "\textit{There's no point in arguing with such extreme attacks. They're just clowns}". In these cases, removal or suspension was considered more appropriate, a judgment many participants voiced explicitly from a bystander standpoint. In other words, while counterspeech offers greater flexibility, it is ill-suited for deeply motivated or extreme hate. Conversely, in mild or borderline cases, counterspeech was not always necessary, and when used, needed to be carefully matched in tone. Hate expressed out of ignorance or self-deprecating humour, if met with overly critical counterspeech, could push users toward opposition and harm the community climate. As P7 explained, "\textit{(If I were that hate speaker,) I might just be expressing my view without strong malice, but Civilbot framed me as if I had bad intentions. I would feel wronged, or uncomfortable.}" Several participants described this reaction while observing others' exchanges, indicating it was a bystander impression rather than only the hate speakers' concern. This suggests the importance of distinguishing between different stages of stigmatization \cite{linkConceptualizingStigma2001}: initial labelling versus entrenched separation motives call for different responses. For users repeatedly engaging in hate, counterspeech alone is unlikely to change their behaviours, and punitive measures may be warranted (P9, P52, P58).

Within the space where counterspeech is considered appropriate, participants also cautioned against "countering every hate speech". Excessive automation could discourage genuine user engagement. Instead, counterspeech was seen as most valuable when directed at high-visibility and potentially disruptive posts, those that often set the tone of discussion and accelerate diffusion. As P22 put it, "\textit{Only high-attention hate speech needs to be countered, because it already exerts a significant influence.}" This emphasis on visibility came from participants speaking as third-party observers, highlighting what bystanders notice. In such cases, counterspeech was perceived not only as a corrective signal superior to deletion but also as a way to neutralize the dominance of hate. Participants further suggested a proactive role for Civilbot at the early-warning stage—for example, predicting when a hateful post is likely to trigger toxic replies or attract huge attention, and intervening before escalation. As P5 proposed, "\textit{Civilbot could be preventive, responding before harmful speech causes damage.}" Prior research on conversational structure for toxicity detection \cite{saveskiStructureToxicConversations2021}, early signals of antisocial behaviour \cite{zhangConversationsGoneAwry2018a}, and the PMCR framework \cite{goyalYouHaveProve2022} offer useful references for identifying intervention opportunities. Building on these approaches, Civilbot could further adopt methods from explainable toxicity detection \cite{liDeModHolisticTool2025} to justify "\textit{why counterspeech is needed here}" as P35 suggested, thereby enhancing credibility for bystanders.

Overall, decisions about whether to counter hinge on two critical factors: the allocation of attention and the risk of conflict. On the one hand, counterspeech may unintentionally draw more traffic to hate speech \cite{cepollaroCounterspeech2023}, underscoring the need to prioritize high-attention or high-risk content. On the other hand, avoiding counterspeech due to possible conflicts risks falling into "negative peace" \cite{jurgensJustComprehensiveStrategy2019}, where harmful ideas spread unchallenged. The desirable balance lies in using a constructive tone and strategic design to transform attacks into dialogue. As P5 reflected, "\textit{When the hateful comment wasn't too emotional and Civilbot also responded in a mild tone, it gave me confidence that the hate speaker's view could actually be changed. It felt like we could have an exchange and reach a warmer, more peaceful outcome.}"

\subsection{Designing Effective Counterspeech: From Reasoning to Style Adaptation}
\label{sec6.2 how}
Drawing on experimental feedback, we propose an operational workflow for counterspeech: motive identification → reasoning analysis → information retrieval → argument construction → strategy selection → style adaptation. This workflow outlines the design space of Civilbot in answering the question of how to counter, raising three core considerations: what to counter, what evidence to use, and how to organize the response. In the organizational dimension, we further discuss strategy selection, stylistic adaptation, and opportunities beyond text.

\subsubsection{Identifying What to Counter: Motives and Weak Points}
\label{sec6.2.1}
Findings in Section \ref{sec5.1.1} underscore the importance of uncovering the motives behind hate speech. Motives not only shape whether a counterspeech is needed but also determine how it should be constructed. For example, when hate stems from cognitive limitations, cognitive strategies—such as prompting critical reflection through questions—can be effective. When driven by negative emotions rooted in personal experience, affective strategies may help elicit more positive affect. If hate is merely attention-seeking, a combination of appropriate negative tone and condemnatory strategies may be more suitable. As P20 observed, different problems demand different approaches: "\textit{If it's ignorance, use rational data; if it's emotional venting, emotion works better.}"

Ignoring these motives risks superficial responses that fail to reach the core, weakening persuasiveness and even making bystanders persuaded by harmful content. In discussions on social issues, participants also emphasized the need for Civilbot to demonstrate sufficient depth of reasoning, without which it would not be taken seriously. As P4 noted, Civilbot's replies sometimes felt "\textit{too vague}", failing to directly address the hate speaker's claims. In short, effective counterspeech must go beyond surface-level wording to engage with the hate speaker's motives, experiences, and reasoning, revealing implicit meanings. This calls for Civilbot to develop deeper interpretive capacities—for instance, handling implicit communication \cite{liangImplicitCommunicationActionable2019} and employing argument schemes \cite{sahaConsolidatingStrategiesCountering2024} to identify hidden assumptions and logical fallacies that can serve as entry points for counterspeech.

Prior work has suggested that personalizing counterspeech with limited user information \cite{dogancGenericPersonalizedInvestigating2023} could further expose motives and enhance persuasion. Yet such personalization raises ethical concerns and technical costs, warranting caution. At the same time, some participants worried that focusing too heavily on the hate speaker's motives might alienate bystanders or targets of hate. As P23 warned, "\textit{The closer you get to the hate speaker's motive, the farther you may be from the other readers.}" Our interpretation is that motive analysis should not be about appeasing the hate speaker but about sharpening the persuasiveness of counterspeech, especially toward bystanders.

Moreover, as P20 emphasized, Civilbot's role is not necessarily to confront hate speakers individually but also to "\textit{assert and maintain community positions}". Counterspeech should therefore aim to resist hate while sustaining constructive discussions on social issues. From this perspective, Civilbot's focus on motives is not merely about "defeating" hate speakers, but about communicating community norms, preserving deliberative climates, and—in certain cases—creating pathways for re-entry of those who move away from hate \cite{hongOutcomeConstrainedLargeLanguage2024}.

\subsubsection{Enhancing Arguments through Information Support}
\label{sec6.2.2}
After analyzing hate speech, Civilbot must determine its arguments and supporting evidence. Results in \ref{sec5.1.3} show that grounding counterspeech in information support was widely seen as promising. Participants expected Civilbot to integrate theoretical knowledge from psychology, sociology, or law, as well as authoritative data and credible news examples, thereby enhancing credibility and significance while helping Civilbot articulate positions and substantiate claims. From a technical standpoint, augmenting counterspeech with external knowledge bases or real-time retrieval was viewed as feasible, aligning with the paradigm of knowledge-guided generation \cite{chungUnderstandingCounterspeechOnline2023}: retrieving relevant information first, then generating enhanced counterspeech. Another pathway is to leverage curated counter-hate datasets (e.g., candidate arguments extracted from online articles \cite{albanyanFindingAuthenticCounterhate2023}) or fine-tune models with high-quality human-authored counterspeech, making Civilbot's output more faithful to authentic user discourse.

Notably, participants already rated Civilbot as moderately credible, likely because it mainly voiced general perspectives rather than relying heavily on factual references. However, as more factual arguments are incorporated, knowledge augmentation will be essential—both to improve persuasiveness and to mitigate hallucination risks \cite{albanyanFindingAuthenticCounterhate2023}. Some participants further stressed that Civilbot should provide facts rather than opinions, enabling viewers to reason independently and thereby reinforcing credibility. This highlights the necessity of complementing counterspeech with factual evidence.

\subsubsection{Choosing Effective Strategies in Context}
\label{sec6.2.3}
From sentence type to tone to strategic intent, results in Section \ref{sec5.2} indicate that context-sensitive strategies are crucial.

For \textbf{sentence type}, questions showed the greatest potential when combined with cognitive strategies: by offering arguments and evidence, they expanded reasoning space and, supported by positive tone, fostered a guiding atmosphere that encouraged critical thinking among both hate speakers and bystanders. Conversely, when paired with negative tone, questions resembled affective strategies of reproach or denunciation. While such forms attracted attention and satisfied some bystanders' sense of justice, their implicit insincerity often suppressed reflection and provoked resistance. Regarding \textbf{tone}, negative expressions were found effective in certain contexts but carried risks. As a non-human agent, Civilbot using negativity could unintentionally normalize hostility or weaken community climate. Emotional countering may therefore be better suited for human users. For Civilbot, a safer design is to prioritize positive or neutral tone while enabling adaptive modulation: in less extreme cases, constructive tone conveys hope for dialogue; in more intense exchanges, restrained negativity or humor-driven benign offense can achieve balance without eroding atmosphere. On \textbf{strategic intent}, results underscored the centrality of cognitive strategies: fact- and reasoning-based counterspeech was key to persuasion and, when paired with positive tone, conveyed objectivity and neutrality. Affective strategies, however, retained supplementary value—empathy-based appeals could ease tension before reasoning took effect, or intensify condemnation after. Together, these findings suggest Civilbot's strategy selection must achieve higher contextual sensitivity, integrating not only the hate speech itself but also broader cues from conversational background, participant interaction, and community climate.

Beyond the strategies synthesized here, broader frameworks are also instructive. For instance, persuasion-oriented approaches \cite{sahaConsolidatingStrategiesCountering2024} include value-based and structure-based strategies. In our framing, argument schemes are incorporated in Section \ref{sec6.2.2} under reasoning, while value-oriented strategies are discussed in Section \ref{sec6.2.4} as part of Civilbot's role design. Other categories such as denouncing or positive tone naturally align with this section's discussion of contextualized choices.

\subsubsection{Adapting Style to Communities and Roles}
\label{sec6.2.4}
Once counterspeech is equipped with clear motives, solid arguments, and strategic combinations, the next challenge is adapting to specific platforms and cultural contexts. Different platforms host distinct user groups and content norms, shaping prevailing values and hate discourses. For instance, a sexist remark toward women might be challenged on Rednote but endorsed on HoopCHINA. Community conventions, expressive styles, and moderation policies also vary. Thus, Civilbot must tailor its counterspeech to the linguistic style and dominant hate topics of each platform, ensuring it integrates with community norms while steering value transmission. More broadly, culture shapes preferences of effective counterspeech: for example, the Chinese tradition of indirectness and metaphor suggests that counterspeech should align with social communication norms to avoid being perceived as biased \cite{munCounterspeakersPerspectivesUnveiling2024}.

Participants further emphasized the potential of role diversification. Civilbot need not remain solely an "enforcer" but could also act as a mediator or knowledge provider. Some even suggested that engaging with high-quality comments might contribute more to constructive debate than directly confronting hate. This implies that Civilbot's role should adapt to situational demands and strategy choices. In our experiment, we deliberately used a neutral avatar and username to avoid over-anthropomorphizing. Yet, profile design (e.g., avatars) could be leveraged for role embodiment, enhancing style and positioning \cite{jungGreatChainAgents2022}, potentially boosting Civilbot's influence and even enabling it to function as an opinion leader. Such role flexibility was noted by several participants specifically from a bystander viewpoint, who valued seeing Civilbot model diverse constructive roles within the community. Role diversification may also balance two competing needs: serious, restrained expression to uphold norms, and more flexible, creative expression to attract attention and foster an anti-hate climate. Future work could explore persona-guided generation, such as configuring dynamic roles through datasets like PersonaChat or leveraging dialogue history for richer role expressions \cite{chungUnderstandingCounterspeechOnline2023}.

\subsubsection{Beyond text: Emerging Modalities}
Currently, only text-based counterspeech is supported by Civilbot, but participants widely envisioned richer modalities. Visualizations, for example, can present arguments more intuitively and capture attention more constructively than negative tones. At the same time, hate is often spread through images \cite{dixonArtisticCounterSpeech2022}, such as malicious memes, posing new detection and response challenges. Integrating multimodal generation—combining text with images—thus represents a promising direction. Beyond text and images, broader design opportunities are opened by interactive digital narratives (IDN) and related formats such as video games and VR/AR/XR \cite{silvaFightingHateSpeech2023}. Anti-hate narratives can be delivered through these media in immersive, participatory environments, extending Civilbot’s reach beyond textual dialogue and community threads into richer experiential domains.

\subsection{Identity in Counterspeech: How It Shapes Chatbot Intervention}
Our findings reveal the impact of Civilbot on bystanders when employing various counterspeech strategies. It is crucial to clarify that this impact stems neither solely from the "strategy itself" nor merely from the fact that it is "chatbot-mediated," but rather from the interplay of both. As noted in Section \ref{sec5.2.2}, bystanders perceive Civilbot differently from humans even when identical tones are used. For instance, when Civilbot adopted a hostile tone, participants (e.g., P5, P9, P24, P26) raised ethical concerns regarding "\textit{machines attacking humans}". P24 explicitly preferred taking personal ownership of intense negative expressions rather than delegating them to Civilbot. Similarly, affective strategies (e.g., "I stand with X") triggered resistance among some participants (P7, P9). Civilbot’s limited agency made such statements prone to being perceived as over-commitment. Moreover, the use of first-person pronouns or empathetic phrasing paradoxically highlighted the agent's non-human identity. Strategies originally intended to bridge psychological distance \cite{zhangYouCompleteMe2022a} instead accentuated identity differences in this sensitive social context, creating a sense of detachment.

These phenomena point to a fundamental issue: the identity of the counterspeaker shapes the intervention's efficacy. This aligns with existing literature suggesting that high-status members or those with clear commitment are more likely to have their counterspeech mimicked \cite{seeringShapingProAntiSocial2017, beneschCounteringDangerousSpeech2014}, and that factors like race and follower count shape a speaker's community influence \cite{mungerTweetmentEffectsTweeted2017}. In our study, where the counterspeaker is an AI chatbot, the implications of identity are more nuanced. \textbf{On the positive side}, Civilbot, as a non-human agent, offers a baseline response free from social and reputational costs, serving as a psychological safety net similar to AI in creative tasks \cite{suhAISocialGlue2021}. This baseline response may encourage bystander engagement by lowering entry thresholds (reflecting the reverse motivation noted in Section \ref{sec5.1.2}) or by serving as an "ice-breaker". Furthermore, as shown in section \ref{sec5.1.3}, Civilbot might de-escalate emotions. While partly due to strategy, section \ref{sec5.1.1} suggests identity played a role: participants (e.g., P30) pointed that people were disinclined to argue with a chatbot, preventing conflict escalation. Additionally, the perception of AI as objective may contribute to this effect. \textbf{On the negative side}, disclosing the bot identity can trigger bias, which varies depending on the perceived level of control \cite{ashktorabEffectsCommunicationDirectionality2021}. For instance, some participants in Section \ref{sec5.1.1} dismissed the AI's output as "\textit{mindless stitching}". This echoes the theoretical distinction that different identities imply different commitments (what it is expected to do) and beliefs (what it is believed capable of doing) \cite{wangAdaptiveHumanAgentTeaming2025}. Consequently, the same expressions may yield divergent effects depending on the speaker's identity.

In summary, these insights suggest two key design implications. First, autonomous counterspeech bots (like Civilbot) and AI systems that assist users in writing counterspeech \cite{munCounterspeakersPerspectivesUnveiling2024} should be treated as distinct design paradigms occupying different social ecological niches. In AI-assisted systems, the speaker remains human; thus, design should focus on collaboration, human-centricity, and even personalization to provide authenticity, emotional support, and empowerment. Conversely, for autonomous bots, the speaker is a non-human agent. Design must therefore prioritize the social perception of the chatbot, managing its role, persona, and strategy selection to optimize its influence on groups like bystanders. Second, the role of identity in counterspeech warrants systematic future research. For example, studies could manipulate perceived identity (perceived as human or bot) alongside conversational style (human or bot), similar to the Ideabot in creativity task \cite{hwangIdeaBotInvestigatingSocial2021}. Furthermore, P23 suggested the possibility of a "\textit{hybrid identity}", where Civilbot’s content is known to be partially authored by humans without explicitly disclosing the source of each message, which may be a meaningful direction of future research.

\subsection{Limitations and Future Directions}
This study employed a lab study to examine Civilbot's impact on bystanders, striving to simulate authentic browsing contexts, yet several limitations remain. 
\begin{itemize}
    \item First, the lab environment may suppress natural behaviours: for instance, P19 noted that the lack of privacy led them to remain silent. While anonymity could theoretically mitigate this, in our study researcher presence was unavoidable due to the need for follow-up interviews based on the interaction process. We sought to compensate by eliciting potential comments through interviews. Future work could explore more covert or automated data collection methods to reduce external interference.
    \item Second, this was a short-term experiment. This design balanced experimental control with participant burden, allowing focus on immediate effects. However, long-term behavioural trends and community dynamics may manifest additional variations. Future studies should conduct longitudinal field study, incorporating natural interactions, such as having hate speech appear in a natural comment sequence, to evaluate Civilbot's sustained impact more comprehensively.
    \item Third, to isolate strategy impact, methods were employed to mitigate context effects, including offering a pool of questions for selection, randomizing question and strategy order, and anonymizing all posted answers. Nevertheless, we recognize that different contexts (e.g., the hate speech, the question topic, and the hate speaker) may necessitate distinct counterspeech strategies, leading to varied effects. As noted in Sections \ref{sec6.1 when} and \ref{sec6.2.3}, the decision to counter and the choice of strategy are linked to the hate speech’s tone, intent, and other factors. Furthermore, Section \ref{sec6.2.1} underscores the importance of discerning the underlying motivations of the hate speech. Future work, therefore, requires a deeper exploration of the compound effect of context $\times$ strategy, ultimately aiming to realize a context-sensitive adaptive mechanism for counterspeech strategies.
    \item Fourth, the study focused on common counter questions, categorizing sentence types as "question" versus "non-question" to highlight the potential of questions in fostering critical thinking. This simplification overlooks potential differences from statements, exclamations, and other sentence forms. Future work could compare sentence types at finer granularity and incorporate role-based strategies such as value to support more nuanced modelling of counterspeech.
    \item Fifth, the study concentrated on the Chinese context. Yet communication norms differ across languages and cultures, which may affect counterspeech reception. Cross-linguistic and cross-cultural studies are needed to assess the generalizability of Civilbot.
    \item Finally, Civilbot relied on pre-generated content rather than real-time detection and generation. While this ensured experimental control, it differs from deployment scenarios. Future research should explore real-time detection, context-aware generation, and integration with platform mechanisms to enhance practical applicability.
\end{itemize}

\section{Ethical Consideration}
During the conduct of this study, we prioritized ethical considerations for all participants. The study protocol was reviewed and approved by the university's Institutional Review Board (IRB), with all required materials submitted according to regulations. Participants were fully briefed on the study's purpose, procedures, potential risks, and their rights, and provided informed consent before participation. They were explicitly informed that participation was voluntary and that they could withdraw at any time without penalty. Psychological support was made available if participants experienced discomfort. All collected data were anonymized, securely stored, and transmitted using encryption to ensure privacy and protect participants' identities. Compensation was provided to participants who completed all sessions.

Additionally, to illustrate Civilbot's counterspeech, we included a small number of real hate‐speech examples with minimal redaction in this paper. These excerpts were selected solely for research transparency and are not intended to perpetuate harmful language. Readers are advised that these examples may contain sensitive content.

\section{Conclusion}
\label{sec7 conclusion}
Counterspeech is proposed as a non-repressive, socially grounded response to online hate incidents, complementing traditional moderation by enriching rather than restricting public discourse. We constructed a unified framework of common counterspeech strategies across sentence type, tone, and strategic intent. Building on this framework, we designed Civilbot, a prototype chatbot capable of generating diverse counterspeech responses, and conducted a mixed-methods, within-subjects experiment to examine its influence on bystanders. Our findings show that Civilbot shaped bystander responses at multiple levels. It was generally perceived as credible and norm-affirming, though its shallow reasoning constrained persuasiveness. Behaviourally, its influence was subtle and sometimes contradictory—providing guidance, substitution, negative modelling, or reverse motivation—while also extending beyond persuasion to fostering community climate. Strategy proved decisive: cognitive strategies outperformed affective ones; tone might influence behavioural tendencies but required contextual calibration; and question forms, when combined with other strategies, could either stimulate critical reflection or provoke resistance. Taken together, these results point to counterspeech design that centres on cognitive strategies while flexibly combining styles in a context-sensitive manner. These findings inform design directions for future counterspeech chatbots. Beyond deciding when to intervene, effective design must also consider how to structure reasoning, integrate credible evidence, and adapt style to community norms. Expanding beyond text into multimodal formats and flexible role configurations can further enhance impact. Ultimately, such chatbots hold potential to mobilize bystanders and cultivate healthier discursive environments against online hate.

\section{Acknowledgments of the Use of AI}
We used AI, in particular large language models (LLMs), in the following ways: (1) Dataset annotation and filtering for hate speech: Qwen-Turbo was used to re-screen entries and tag targeted groups. (2) Counterspeech generation: GPT-5 was employed to generate Civilbot responses according to the eight predefined counterspeech strategies, using iterative prompt engineering to refine outputs. Details of these usages are provided in Section \ref{sec4 experiment design} and Appendix \ref{app:prompt}. Additionally, the full manuscript was AI-assisted for language polishing and grammar checking only; all content decisions, analyses, and interpretations were made by the authors. Authors take full responsibility for the output and use of AI in this paper.

\begin{acks}
This research was supported by National Natural Science Foundation of China (NSFC) under the Grant No. 62372113. Peng Zhang is a faculty of College of Computer Science and Artificial Intelligence, Fudan University. Tun Lu is a faculty of College of Computer Science and Artificial Intelligence, Shanghai Key Laboratory of Data Science, and Silver-X MOE Philosophy \& Social Sciences Laboratory, Fudan Institute on Aging.
\end{acks}

\bibliographystyle{ACM-Reference-Format}

\bibliography{base}

@article{khokharTheoryDevelopmentThematic2020,
  title = {Theory {{Development}} in {{Thematic Analysis}}: {{Procedure}} and {{Practice}}},
  shorttitle = {Theory {{Development}} in {{Thematic Analysis}}},
  author = {Khokhar, Sameena and Pathan, Habibullah and Raheem, Arsalan and Abbasi, Abdul Malik},
  date = {2020-12-31},
  year = {2020},
  journaltitle = {Review of Applied Management and Social Sciences},
  shortjournal = {RAMSS},
  volume = {3},
  number = {3},
  pages = {423--433},
  issn = {2708-3640, 2708-2024},
  doi = {10.47067/ramss.v3i3.79},
  url = {http://ramss.spcrd.org/index.php/ramss/article/view/79},
  urldate = {2025-09-08},
  langid = {american}
}

@book{girdenANOVA1992,
  title = {{{ANOVA}}},
  author = {Girden, Ellen},
  date = {1992},
  year = {1992},
  publisher = {SAGE Publications, Inc.},
  location = {2455 Teller Road,~Thousand Oaks~California~91320~United States of America},
  doi = {10.4135/9781412983419},
  url = {https://methods.sagepub.com/book/anova},
  urldate = {2025-09-08},
  isbn = {978-0-8039-4257-8 978-1-4129-8341-9}
}

@article{fleissMeasuringNominalScale1971,
  title = {Measuring Nominal Scale Agreement among Many Raters.},
  author = {Fleiss, Joseph L.},
  date = {1971-11},
  year = {1971},
  journaltitle = {Psychological Bulletin},
  shortjournal = {Psychological Bulletin},
  volume = {76},
  number = {5},
  pages = {378--382},
  issn = {1939-1455, 0033-2909},
  doi = {10.1037/h0031619},
  url = {https://doi.apa.org/doi/10.1037/h0031619},
  urldate = {2025-09-09},
  langid = {english}
}

@inproceedings{albanyanFindingAuthenticCounterhate2023,
  title = {Finding {{Authentic Counterhate Arguments}}: {{A Case Study}} with {{Public Figures}}},
  shorttitle = {Finding {{Authentic Counterhate Arguments}}},
  booktitle = {Proceedings of the 2023 {{Conference}} on {{Empirical Methods}} in {{Natural Language Processing}}},
  author = {Albanyan, Abdullah and Hassan, Ahmed and Blanco, Eduardo},
  editor = {Bouamor, Houda and Pino, Juan and Bali, Kalika},
  year = {2023},
  month = dec,
  pages = {13862--13876},
  publisher = {Association for Computational Linguistics},
  address = {Singapore},
  doi = {10.18653/v1/2023.emnlp-main.855},
  urldate = {2025-04-25},
  langid = {american}
}

@misc{aleksandricSadnessAngerAnxiety2023,
  title = {Sadness, {{Anger}}, or {{Anxiety}}: {{Twitter Users}}' {{Emotional Responses}} to {{Toxicity}} in {{Public Conversations}}},
  shorttitle = {Sadness, {{Anger}}, or {{Anxiety}}},
  author = {Aleksandric, Ana and Pankaj, Hanani and Wilson, Gabriela Mustata and Nilizadeh, Shirin},
  year = {2023},
  month = oct,
  number = {arXiv:2310.11436},
  eprint = {2310.11436},
  publisher = {arXiv},
  doi = {10.48550/arXiv.2310.11436},
  urldate = {2025-04-21},
  archiveprefix = {arXiv},
  langid = {american}
}

@inproceedings{aleksandricUsersBehavioralEmotional2024,
  title = {Users' {{Behavioral}} and {{Emotional Response}} to {{Toxicity}} in {{Twitter Conversations}}},
  booktitle = {Proceedings of the {{International AAAI Conference}} on {{Web}} and {{Social Media}}},
  author = {Aleksandric, Ana and Roy, Sayak Saha and Pankaj, Hanani and Wilson, Gabriela Mustata and Nilizadeh, Shirin},
  year = {2024},
  volume = {18},
  pages = {29--42},
  urldate = {2025-04-21},
  langid = {american}
}

@inproceedings{ashktorabEffectsCommunicationDirectionality2021,
  title = {Effects of {{Communication Directionality}} and {{AI Agent Differences}} in {{Human-AI Interaction}}},
  booktitle = {Proceedings of the 2021 {{CHI Conference}} on {{Human Factors}} in {{Computing Systems}}},
  author = {Ashktorab, Zahra and Dugan, Casey and Johnson, James and Pan, Qian and Zhang, Wei and Kumaravel, Sadhana and Campbell, Murray},
  year = {2021},
  month = may,
  series = {{{CHI}} '21},
  pages = {1--15},
  publisher = {Association for Computing Machinery},
  address = {New York, NY, USA},
  doi = {10.1145/3411764.3445256},
  urldate = {2024-08-06},
  isbn = {978-1-4503-8096-6}
}

@article{baddeleyHerdingSocialInfluence2010,
  title = {Herding, Social Influence and Economic Decision-Making: Socio-Psychological and Neuroscientific Analyses},
  shorttitle = {Herding, Social Influence and Economic Decision-Making},
  author = {Baddeley, Michelle},
  year = {2010},
  month = jan,
  journal = {Philosophical Transactions of the Royal Society B: Biological Sciences},
  volume = {365},
  number = {1538},
  pages = {281--290},
  issn = {0962-8436, 1471-2970},
  doi = {10.1098/rstb.2009.0169},
  urldate = {2025-09-02},
  langid = {english}
}

@misc{barGenerativeAIMay2024,
  title = {Generative {{AI}} May Backfire for Counterspeech},
  author = {B{\"a}r, Dominik and Maarouf, Abdurahman and Feuerriegel, Stefan},
  year = {2024},
  month = nov,
  number = {arXiv:2411.14986},
  eprint = {2411.14986},
  publisher = {arXiv},
  doi = {10.48550/arXiv.2411.14986},
  urldate = {2025-05-07},
  archiveprefix = {arXiv},
  langid = {american}
}

@misc{beneschCounteringDangerousSpeech2014,
  type = {{{SSRN Scholarly Paper}}},
  title = {Countering {{Dangerous Speech}}: {{New Ideas}} for {{Genocide Prevention}}},
  shorttitle = {Countering {{Dangerous Speech}}},
  author = {Benesch, Susan},
  year = {2014},
  month = feb,
  number = {3686876},
  eprint = {3686876},
  publisher = {Social Science Research Network},
  address = {Rochester, NY},
  doi = {10.2139/ssrn.3686876},
  urldate = {2025-05-14},
  archiveprefix = {Social Science Research Network},
  langid = {english}
}

@misc{benniePANDAPairedAntihate2025,
  title = {{{PANDA}} -- {{Paired Anti-hate Narratives Dataset}} from {{Asia}}: {{Using}} an {{LLM-as-a-Judge}} to {{Create}} the {{First Chinese Counterspeech Dataset}}},
  shorttitle = {{{PANDA}} -- {{Paired Anti-hate Narratives Dataset}} from {{Asia}}},
  author = {Bennie, Michael and Zhang, Demi and Xiao, Bushi and Cao, Jing and Liu, Chryseis Xinyi and Meng, Jian and Tripp, Alayo},
  year = {2025},
  month = jan,
  number = {arXiv:2501.00697},
  eprint = {2501.00697},
  publisher = {arXiv},
  doi = {10.48550/arXiv.2501.00697},
  urldate = {2025-06-16},
  archiveprefix = {arXiv},
  langid = {american}
}

@inproceedings{berryDiscussionQualityDiffuses2017,
  title = {Discussion {{Quality Diffuses}} in the {{Digital Public Square}}},
  booktitle = {Proceedings of the 26th {{International Conference}} on {{World Wide Web}}},
  author = {Berry, George and Taylor, Sean J.},
  year = {2017},
  month = apr,
  pages = {1371--1380},
  publisher = {International World Wide Web Conferences Steering Committee},
  address = {Perth Australia},
  doi = {10.1145/3038912.3052666},
  urldate = {2025-05-13},
  isbn = {978-1-4503-4913-0},
  langid = {english}
}

@article{bilewiczArtificialIntelligenceHate2021,
  title = {Artificial Intelligence against Hate: {{Intervention}} Reducing Verbal Aggression in the Social Network Environment},
  shorttitle = {Artificial Intelligence against Hate},
  author = {Bilewicz, Micha{\l} and Tempska, Patrycja and Leliwa, Gniewosz and Dowgia{\l}{\l}o, Maria and Ta{\'n}ska, Michalina and Urbaniak, Rafa{\l} and Wroczy{\'n}ski, Micha{\l}},
  year = {2021},
  month = may,
  journal = {Aggressive Behavior},
  volume = {47},
  number = {3},
  pages = {260--266},
  issn = {0096-140X, 1098-2337},
  doi = {10.1002/ab.21948},
  urldate = {2025-05-09},
  langid = {english}
}

@inproceedings{bonaldiNLPCounterspeechHate2024,
  title = {{{NLP}} for {{Counterspeech}} against {{Hate}}: {{A Survey}} and {{How-To Guide}}},
  shorttitle = {{{NLP}} for {{Counterspeech}} against {{Hate}}},
  booktitle = {Findings of the {{Association}} for {{Computational Linguistics}}: {{NAACL}} 2024},
  author = {Bonaldi, Helena and Chung, Yi-Ling and Abercrombie, Gavin and Guerini, Marco},
  editor = {Duh, Kevin and Gomez, Helena and Bethard, Steven},
  year = {2024},
  month = jun,
  pages = {3480--3499},
  publisher = {Association for Computational Linguistics},
  address = {Mexico City, Mexico},
  doi = {10.18653/v1/2024.findings-naacl.221},
  urldate = {2025-04-21},
  langid = {american}
}

@incollection{bromellCounterSpeechEveryonesResponsibility2022,
  title = {Counter-{{Speech Is Everyone}}'s {{Responsibility}}},
  booktitle = {Regulating {{Free Speech}} in a {{Digital Age}}: {{Hate}}, {{Harm}} and the {{Limits}} of {{Censorship}}},
  author = {Bromell, David},
  editor = {Bromell, David},
  year = {2022},
  pages = {191--215},
  publisher = {Springer International Publishing},
  address = {Cham},
  urldate = {2025-05-07},
  isbn = {978-3-030-95550-2},
  langid = {english}
}

@article{cepollaroCounterspeech2023,
  title = {Counterspeech},
  author = {Cepollaro, Bianca and Lepoutre, Maxime and Simpson, Robert Mark},
  year = {2023},
  journal = {Philosophy Compass},
  volume = {18},
  number = {1},
  pages = {e12890},
  issn = {1747-9991},
  doi = {10.1111/phc3.12890},
  urldate = {2025-05-07},
  copyright = {{\copyright} 2022 The Authors. Philosophy Compass published by John Wiley \& Sons Ltd.},
  langid = {english}
}

@inproceedings{chengAnyoneCanBecome2017,
  title = {Anyone {{Can Become}} a {{Troll}}: {{Causes}} of {{Trolling Behavior}} in {{Online Discussions}}},
  shorttitle = {Anyone {{Can Become}} a {{Troll}}},
  booktitle = {Proceedings of the 2017 {{ACM Conference}} on {{Computer Supported Cooperative Work}} and {{Social Computing}}},
  author = {Cheng, Justin and Bernstein, Michael and {Danescu-Niculescu-Mizil}, Cristian and Leskovec, Jure},
  year = {2017},
  month = feb,
  series = {{{CSCW}} '17},
  pages = {1217--1230},
  publisher = {Association for Computing Machinery},
  address = {New York, NY, USA},
  doi = {10.1145/2998181.2998213},
  urldate = {2025-04-17},
  isbn = {978-1-4503-4335-0},
  langid = {american}
}

@inproceedings{chungCONANCOunterNArratives2019,
  title = {{{CONAN}} - {{COunter NArratives}} through {{Nichesourcing}}: A {{Multilingual Dataset}} of {{Responses}} to {{Fight Online Hate Speech}}},
  shorttitle = {{{CONAN}} - {{COunter NArratives}} through {{Nichesourcing}}},
  booktitle = {Proceedings of the 57th {{Annual Meeting}} of the {{Association}} for {{Computational Linguistics}}},
  author = {Chung, Yi-Ling and Kuzmenko, Elizaveta and Tekiroglu, Serra Sinem and Guerini, Marco},
  editor = {Korhonen, Anna and Traum, David and M{\`a}rquez, Llu{\'i}s},
  year = {2019},
  month = jul,
  pages = {2819--2829},
  publisher = {Association for Computational Linguistics},
  address = {Florence, Italy},
  doi = {10.18653/v1/P19-1271},
  urldate = {2025-06-04},
  langid = {american}
}

@misc{chungUnderstandingCounterspeechOnline2023,
  title = {Understanding {{Counterspeech}} for {{Online Harm Mitigation}}},
  author = {Chung, Yi-Ling and Abercrombie, Gavin and Enock, Florence and Bright, Jonathan and Rieser, Verena},
  year = {2023},
  month = jul,
  number = {arXiv:2307.04761},
  eprint = {2307.04761},
  publisher = {arXiv},
  doi = {10.48550/arXiv.2307.04761},
  urldate = {2025-05-07},
  archiveprefix = {arXiv},
  langid = {american}
}

@inproceedings{cimaContextualizedCounterspeechStrategies2025a,
  title = {Contextualized {{Counterspeech}}: {{Strategies}} for {{Adaptation}}, {{Personalization}}, and {{Evaluation}}},
  shorttitle = {Contextualized {{Counterspeech}}},
  booktitle = {Proceedings of the {{ACM}} on {{Web Conference}} 2025},
  author = {Cima, Lorenzo and Miaschi, Alessio and Trujillo, Amaury and Avvenuti, Marco and Dell'Orletta, Felice and Cresci, Stefano},
  year = {2025},
  month = apr,
  series = {{{WWW}} '25},
  pages = {5022--5033},
  publisher = {Association for Computing Machinery},
  address = {New York, NY, USA},
  doi = {10.1145/3696410.3714507},
  urldate = {2025-05-07},
  isbn = {979-8-4007-1274-6},
  langid = {american}
}

@article{cohenStatisticalPowerAnalysis1992,
  title = {Statistical {{Power Analysis}}},
  author = {Cohen, Jacob},
  year = {1992},
  journal = {Current Directions in Psychological Science},
  volume = {1},
  number = {3},
  eprint = {20182143},
  eprinttype = {jstor},
  pages = {98--101},
  issn = {0963-7214},
  urldate = {2025-07-31}
}

@inproceedings{danryDontJustTell2023b,
  title = {Don't {{Just Tell Me}}, {{Ask Me}}: {{AI Systems}} That {{Intelligently Frame Explanations}} as {{Questions Improve Human Logical Discernment Accuracy}} over {{Causal AI}} Explanations},
  shorttitle = {Don't {{Just Tell Me}}, {{Ask Me}}},
  booktitle = {Proceedings of the 2023 {{CHI Conference}} on {{Human Factors}} in {{Computing Systems}}},
  author = {Danry, Valdemar and Pataranutaporn, Pat and Mao, Yaoli and Maes, Pattie},
  year = {2023},
  month = apr,
  series = {{{CHI}} '23},
  pages = {1--13},
  publisher = {Association for Computing Machinery},
  address = {New York, NY, USA},
  doi = {10.1145/3544548.3580672},
  urldate = {2024-08-07},
  isbn = {978-1-4503-9421-5},
  langid = {american}
}

@misc{dingCounterQuillInvestigatingPotential2025,
  title = {{{CounterQuill}}: {{Investigating}} the {{Potential}} of {{Human-AI Collaboration}} in {{Online Counterspeech Writing}}},
  shorttitle = {{{CounterQuill}}},
  author = {Ding, Xiaohan and Ping, Kaike and Gunturi, Uma Sushmitha and Carik, Buse and Stil, Sophia and Wilhelm, Lance T. and Daryanto, Taufiq and Hawdon, James and Lee, Sang Won and Rho, Eugenia H.},
  year = {2025},
  month = may,
  number = {arXiv:2410.03032},
  eprint = {2410.03032},
  publisher = {arXiv},
  doi = {10.48550/arXiv.2410.03032},
  urldate = {2025-05-07},
  archiveprefix = {arXiv},
  langid = {american}
}

@article{dixonArtisticCounterSpeech2022,
  title = {Artistic ({{Counter}}) {{Speech}}},
  author = {Dixon, Daisy},
  year = {2022},
  month = sep,
  journal = {The Journal of Aesthetics and Art Criticism},
  volume = {80},
  number = {4},
  pages = {409--419},
  issn = {0021-8529},
  doi = {10.1093/jaac/kpac038},
  urldate = {2025-05-07},
  langid = {american}
}

@inproceedings{dogancGenericPersonalizedInvestigating2023,
  title = {From {{Generic}} to {{Personalized}}: {{Investigating Strategies}} for {{Generating Targeted Counter Narratives}} against {{Hate Speech}}},
  shorttitle = {From {{Generic}} to {{Personalized}}},
  booktitle = {Proceedings of the 1st {{Workshop}} on {{CounterSpeech}} for {{Online Abuse}} ({{CS4OA}})},
  author = {Do{\u g}an{\c c}, Mekselina and Markov, Ilia},
  editor = {Chung, Yi-Ling and Bonaldi, Helena and Abercrombie, Gavin and Guerini, Marco},
  year = {2023},
  month = sep,
  pages = {1--12},
  publisher = {Association for Computational Linguistics},
  address = {Prague, Czechia},
  urldate = {2025-05-07},
  langid = {american}
}

@inproceedings{goversAIDrivenMediationStrategies2024,
  title = {{{AI-Driven Mediation Strategies}} for {{Audience Depolarisation}} in {{Online Debates}}},
  booktitle = {Proceedings of the 2024 {{CHI Conference}} on {{Human Factors}} in {{Computing Systems}}},
  author = {Govers, Jarod and Velloso, Eduardo and Kostakos, Vassilis and Goncalves, Jorge},
  year = {2024},
  month = may,
  series = {{{CHI}} '24},
  pages = {1--18},
  publisher = {Association for Computing Machinery},
  address = {New York, NY, USA},
  doi = {10.1145/3613904.3642322},
  urldate = {2025-03-31},
  isbn = {979-8-4007-0330-0},
  langid = {american}
}

@inproceedings{goyalYouHaveProve2022,
  title = {''{{You}} Have to Prove the Threat Is Real'': {{Understanding}} the Needs of {{Female Journalists}} and {{Activists}} to {{Document}} and {{Report Online Harassment}}},
  shorttitle = {''{{You}} Have to Prove the Threat Is Real''},
  booktitle = {Proceedings of the 2022 {{CHI Conference}} on {{Human Factors}} in {{Computing Systems}}},
  author = {Goyal, Nitesh and Park, Leslie and Vasserman, Lucy},
  year = {2022},
  month = apr,
  series = {{{CHI}} '22},
  pages = {1--17},
  publisher = {Association for Computing Machinery},
  address = {New York, NY, USA},
  doi = {10.1145/3491102.3517517},
  urldate = {2025-05-15},
  isbn = {978-1-4503-9157-3},
  langid = {american}
}

@inproceedings{guptaCounterspeechesMySleeve2023,
  title = {Counterspeeches up My Sleeve! {{Intent Distribution Learning}} and {{Persistent Fusion}} for {{Intent-Conditioned Counterspeech Generation}}},
  booktitle = {Proceedings of the 61st {{Annual Meeting}} of the {{Association}} for {{Computational Linguistics}} ({{Volume}} 1: {{Long Papers}})},
  author = {Gupta, Rishabh and Desai, Shaily and Goel, Manvi and Bandhakavi, Anil and Chakraborty, Tanmoy and Akhtar, Md. Shad},
  editor = {Rogers, Anna and {Boyd-Graber}, Jordan and Okazaki, Naoaki},
  year = {2023},
  month = jul,
  pages = {5792--5809},
  publisher = {Association for Computational Linguistics},
  address = {Toronto, Canada},
  doi = {10.18653/v1/2023.acl-long.318},
  urldate = {2025-04-25},
  langid = {american}
}

@inproceedings{halimWokeGPTImprovingCounterspeech2023a,
  title = {{{WokeGPT}}: {{Improving Counterspeech Generation Against Online Hate Speech}} by {{Intelligently Augmenting Datasets Using}} a {{Novel Metric}}},
  shorttitle = {{{WokeGPT}}},
  booktitle = {2023 {{International Joint Conference}} on {{Neural Networks}} ({{IJCNN}})},
  author = {Halim, Sadaf MD and Irtiza, Saquib and Hu, Yibo and Khan, Latifur and Thuraisingham, Bhavani},
  year = {2023},
  month = jun,
  pages = {1--10},
  doi = {10.1109/IJCNN54540.2023.10191114},
  urldate = {2025-05-07},
  langid = {american}
}

@article{hangartnerEmpathybasedCounterspeechCan2021,
  title = {Empathy-Based Counterspeech Can Reduce Racist Hate Speech in a Social Media Field Experiment},
  author = {Hangartner, Dominik and Gennaro, Gloria and Alasiri, Sary and Bahrich, Nicholas and Bornhoft, Alexandra and Boucher, Joseph and Demirci, Buket Buse and Derksen, Laurenz and Hall, Aldo and Jochum, Matthias and Munoz, Maria Murias and Richter, Marc and Vogel, Franziska and Wittwer, Salom{\'e} and W{\"u}thrich, Felix and Gilardi, Fabrizio and Donnay, Karsten},
  year = {2021},
  month = dec,
  journal = {Proceedings of the National Academy of Sciences},
  volume = {118},
  number = {50},
  pages = {e2116310118},
  doi = {10.1073/pnas.2116310118},
  urldate = {2025-05-07},
  langid = {american}
}

@article{hanPlayingNiceModeling2015,
  title = {Playing {{Nice}}: {{Modeling Civility}} in {{Online Political Discussions}}},
  shorttitle = {Playing {{Nice}}},
  author = {Han, Soo-Hye and Brazeal, LeAnn M.},
  year = {2015},
  month = jan,
  journal = {Communication Research Reports},
  volume = {32},
  number = {1},
  pages = {20--28},
  issn = {0882-4096, 1746-4099},
  doi = {10.1080/08824096.2014.989971},
  urldate = {2025-05-13},
  langid = {english}
}

@inproceedings{hartmannLostModerationHow2025,
  title = {Lost in {{Moderation}}: {{How Commercial Content Moderation APIs Over-}} and {{Under-Moderate Group-Targeted Hate Speech}} and {{Linguistic Variations}}},
  shorttitle = {Lost in {{Moderation}}},
  booktitle = {Proceedings of the 2025 {{CHI Conference}} on {{Human Factors}} in {{Computing Systems}}},
  author = {Hartmann, David and Oueslati, Amin and Staufer, Dimitri and Pohlmann, Lena and Munzert, Simon and Heuer, Hendrik},
  year = {2025},
  month = apr,
  series = {{{CHI}} '25},
  pages = {1--26},
  publisher = {Association for Computing Machinery},
  address = {New York, NY, USA},
  doi = {10.1145/3706598.3713998},
  urldate = {2025-05-15},
  isbn = {979-8-4007-1394-1},
  langid = {american}
}

@inproceedings{heRacismVirusAntiasian2022,
  title = {Racism Is a Virus: Anti-Asian Hate and Counterspeech in Social Media during the {{COVID-19}} Crisis},
  shorttitle = {Racism Is a Virus},
  booktitle = {Proceedings of the 2021 {{IEEE}}/{{ACM International Conference}} on {{Advances}} in {{Social Networks Analysis}} and {{Mining}}},
  author = {He, Bing and Ziems, Caleb and Soni, Sandeep and Ramakrishnan, Naren and Yang, Diyi and Kumar, Srijan},
  year = {2022},
  month = jan,
  series = {{{ASONAM}} '21},
  pages = {90--94},
  publisher = {Association for Computing Machinery},
  address = {New York, NY, USA},
  doi = {10.1145/3487351.3488324},
  urldate = {2025-05-14},
  isbn = {978-1-4503-9128-3},
  langid = {american}
}

@misc{hongOutcomeConstrainedLargeLanguage2024,
  title = {Outcome-{{Constrained Large Language Models}} for {{Countering Hate Speech}}},
  author = {Hong, Lingzi and Luo, Pengcheng and Blanco, Eduardo and Song, Xiaoying},
  year = {2024},
  month = oct,
  number = {arXiv:2403.17146},
  eprint = {2403.17146},
  publisher = {arXiv},
  doi = {10.48550/arXiv.2403.17146},
  urldate = {2025-06-16},
  archiveprefix = {arXiv},
  langid = {american}
}

@inproceedings{hwangIdeaBotInvestigatingSocial2021,
  title = {{{IdeaBot}}: {{Investigating Social Facilitation}} in {{Human-Machine Team Creativity}}},
  shorttitle = {{{IdeaBot}}},
  booktitle = {Proceedings of the 2021 {{CHI Conference}} on {{Human Factors}} in {{Computing Systems}}},
  author = {Hwang, Angel Hsing-Chi and Won, Andrea Stevenson},
  year = {2021},
  month = may,
  series = {{{CHI}} '21},
  pages = {1--16},
  publisher = {Association for Computing Machinery},
  address = {New York, NY, USA},
  doi = {10.1145/3411764.3445270},
  urldate = {2024-08-06},
  isbn = {978-1-4503-8096-6},
  langid = {american}
}

@article{jiaTacklingHateSpeech2025,
  title = {Tackling Hate Speech Online: {{The}} Effect of Counter-Speech on Subsequent Bystander Behavioral Intentions},
  shorttitle = {Tackling Hate Speech Online},
  author = {Jia, Yue and Schumann, Sandy},
  year = {2025},
  journal = {Cyberpsychology: Journal of Psychosocial Research on Cyberspace},
  volume = {19},
  number = {1},
  urldate = {2025-05-07},
  langid = {american}
}

@misc{jurgensJustComprehensiveStrategy2019,
  title = {A {{Just}} and {{Comprehensive Strategy}} for {{Using NLP}} to {{Address Online Abuse}}},
  author = {Jurgens, David and Chandrasekharan, Eshwar and Hemphill, Libby},
  year = {2019},
  month = jun,
  number = {arXiv:1906.01738},
  eprint = {1906.01738},
  publisher = {arXiv},
  doi = {10.48550/arXiv.1906.01738},
  urldate = {2025-04-24},
  archiveprefix = {arXiv},
  langid = {american}
}

@article{kramerExperimentalEvidenceMassivescale2014,
  title = {Experimental Evidence of Massive-Scale Emotional Contagion through Social Networks},
  author = {Kramer, Adam D. I. and Guillory, Jamie E. and Hancock, Jeffrey T.},
  year = {2014},
  month = jun,
  journal = {Proceedings of the National Academy of Sciences},
  volume = {111},
  number = {24},
  pages = {8788--8790},
  issn = {0027-8424, 1091-6490},
  doi = {10.1073/pnas.1320040111},
  urldate = {2025-08-14},
  langid = {english}
}

@article{langtonBlockingCounterspeech2018,
  title = {Blocking as Counter-Speech},
  author = {Langton, Rae},
  year = {2018},
  journal = {New work on speech acts},
  volume = {144},
  pages = {156},
  urldate = {2025-05-07},
  langid = {american}
}

@book{lepoutreDemocraticSpeechDivided2021,
  title = {Democratic Speech in Divided Times},
  author = {Lepoutre, Maxime},
  year = {2021},
  publisher = {Oxford University Press},
  urldate = {2025-05-29},
  langid = {american}
}

@inproceedings{liangImplicitCommunicationActionable2019,
  title = {Implicit {{Communication}} of {{Actionable Information}} in {{Human-AI}} Teams},
  booktitle = {Proceedings of the 2019 {{CHI Conference}} on {{Human Factors}} in {{Computing Systems}}},
  author = {Liang, Claire and Proft, Julia and Andersen, Erik and Knepper, Ross A.},
  year = {2019},
  month = may,
  series = {{{CHI}} '19},
  pages = {1--13},
  publisher = {Association for Computing Machinery},
  address = {New York, NY, USA},
  doi = {10.1145/3290605.3300325},
  urldate = {2024-08-06},
  isbn = {978-1-4503-5970-2}
}

@article{linkConceptualizingStigma2001,
  title = {Conceptualizing {{Stigma}}},
  author = {Link, Bruce G. and Phelan, Jo C.},
  year = {2001},
  month = aug,
  journal = {Annual Review of Sociology},
  volume = {27},
  number = {1},
  pages = {363--385},
  issn = {0360-0572},
  doi = {10.1146/annurev.soc.27.1.363},
  urldate = {2023-12-28},
  langid = {american}
}

@misc{mathewAnalyzingHateCounter2018,
  title = {Analyzing the Hate and Counter Speech Accounts on {{Twitter}}},
  author = {Mathew, Binny and Kumar, Navish and Ravina and Goyal, Pawan and Mukherjee, Animesh},
  year = {2018},
  month = dec,
  number = {arXiv:1812.02712},
  eprint = {1812.02712},
  publisher = {arXiv},
  doi = {10.48550/arXiv.1812.02712},
  urldate = {2025-05-07},
  archiveprefix = {arXiv},
  langid = {american}
}

@article{mathewThouShaltNot2019,
  title = {Thou {{Shalt Not Hate}}: {{Countering Online Hate Speech}}},
  shorttitle = {Thou {{Shalt Not Hate}}},
  author = {Mathew, Binny and Saha, Punyajoy and Tharad, Hardik and Rajgaria, Subham and Singhania, Prajwal and Maity, Suman Kalyan and Goyal, Pawan and Mukherjee, Animesh},
  year = {2019},
  month = jul,
  journal = {Proceedings of the International AAAI Conference on Web and Social Media},
  volume = {13},
  pages = {369--380},
  issn = {2334-0770},
  doi = {10.1609/icwsm.v13i01.3237},
  urldate = {2025-06-04},
  copyright = {Copyright (c) 2019 Association for the Advancement of Artificial Intelligence},
  langid = {english}
}

@article{molinaRoleCivilityMetacommunication2018,
  title = {The {{Role}} of {{Civility}} and {{Metacommunication}} in {{Facebook Discussions}}},
  author = {Molina, Roc{\'i}o Galarza and Jennings, Freddie J.},
  year = {2018},
  month = jan,
  journal = {Communication Studies},
  volume = {69},
  number = {1},
  pages = {42--66},
  issn = {1051-0974, 1745-1035},
  doi = {10.1080/10510974.2017.1397038},
  urldate = {2025-05-13},
  langid = {english}
}

@inproceedings{munCounterspeakersPerspectivesUnveiling2024,
  title = {Counterspeakers' {{Perspectives}}: {{Unveiling Barriers}} and {{AI Needs}} in the {{Fight}} against {{Online Hate}}},
  shorttitle = {Counterspeakers' {{Perspectives}}},
  booktitle = {Proceedings of the 2024 {{CHI Conference}} on {{Human Factors}} in {{Computing Systems}}},
  author = {Mun, Jimin and Buerger, Cathy and Liang, Jenny T and Garland, Joshua and Sap, Maarten},
  year = {2024},
  month = may,
  series = {{{CHI}} '24},
  pages = {1--22},
  publisher = {Association for Computing Machinery},
  address = {New York, NY, USA},
  doi = {10.1145/3613904.3642025},
  urldate = {2025-04-17},
  isbn = {979-8-4007-0330-0},
  langid = {american}
}

@article{mungerTweetmentEffectsTweeted2017,
  title = {Tweetment {{Effects}} on the {{Tweeted}}: {{Experimentally Reducing Racist Harassment}}},
  shorttitle = {Tweetment {{Effects}} on the {{Tweeted}}},
  author = {Munger, Kevin},
  year = {2017},
  month = sep,
  journal = {Political Behavior},
  volume = {39},
  number = {3},
  pages = {629--649},
  issn = {0190-9320, 1573-6687},
  doi = {10.1007/s11109-016-9373-5},
  urldate = {2025-08-14},
  langid = {english}
}

@article{noelle-neumannSpiralSilenceTheory1974,
  title = {The Spiral of Silence a Theory of Public Opinion},
  author = {{Noelle-Neumann}, Elisabeth},
  year = {1974},
  journal = {Journal of communication},
  volume = {24},
  number = {2},
  pages = {43--51},
  urldate = {2025-09-02},
  langid = {american}
}

@inbook{pettyElaborationLikelihoodModel1986,
  title = {The {{Elaboration Likelihood Model}} of {{Persuasion}}},
  booktitle = {Communication and {{Persuasion}}},
  author = {Petty, Richard E. and Cacioppo, John T.},
  year = {1986},
  pages = {1--24},
  publisher = {Springer New York},
  address = {New York, NY},
  urldate = {2025-06-03},
  collaborator = {Petty, Richard E. and Cacioppo, John T.},
  isbn = {978-1-4612-9378-1 978-1-4612-4964-1},
  langid = {english}
}

@misc{pingCounterExploringMotivations2024,
  title = {Behind the {{Counter}}: {{Exploring}} the {{Motivations}} and {{Barriers}} of {{Online Counterspeech Writing}}},
  shorttitle = {Behind the {{Counter}}},
  author = {Ping, Kaike and Kumar, Anisha and Ding, Xiaohan and Rho, Eugenia},
  year = {2024},
  month = mar,
  number = {arXiv:2403.17116},
  eprint = {2403.17116},
  publisher = {arXiv},
  doi = {10.48550/arXiv.2403.17116},
  urldate = {2025-05-07},
  archiveprefix = {arXiv},
  langid = {american}
}

@article{pingPerceivingCounteringHate2025,
  title = {Perceiving and {{Countering Hate}}: {{The Role}} of {{Identity}} in {{Online Responses}}},
  shorttitle = {Perceiving and {{Countering Hate}}},
  author = {Ping, Kaike and Hawdon, James and Rho, Eugenia H},
  year = {2025},
  month = may,
  journal = {Proc. ACM Hum.-Comput. Interact.},
  volume = {9},
  number = {2},
  pages = {CSCW147:1--CSCW147:28},
  doi = {10.1145/3711045},
  urldate = {2025-05-16},
  langid = {american}
}

@inproceedings{qianBenchmarkDatasetLearning2019,
  title = {A {{Benchmark Dataset}} for {{Learning}} to {{Intervene}} in {{Online Hate Speech}}},
  booktitle = {Proceedings of the 2019 {{Conference}} on {{Empirical Methods}} in {{Natural Language Processing}} and the 9th {{International Joint Conference}} on {{Natural Language Processing}} ({{EMNLP-IJCNLP}})},
  author = {Qian, Jing and Bethke, Anna and Liu, Yinyin and Belding, Elizabeth and Wang, William Yang},
  editor = {Inui, Kentaro and Jiang, Jing and Ng, Vincent and Wan, Xiaojun},
  year = {2019},
  month = nov,
  pages = {4755--4764},
  publisher = {Association for Computational Linguistics},
  address = {Hong Kong, China},
  doi = {10.18653/v1/D19-1482},
  urldate = {2025-05-14},
  langid = {american}
}

@inproceedings{qianBenchmarkDatasetLearning2019a,
  title = {A {{Benchmark Dataset}} for {{Learning}} to {{Intervene}} in {{Online Hate Speech}}},
  booktitle = {Proceedings of the 2019 {{Conference}} on {{Empirical Methods}} in {{Natural Language Processing}} and the 9th {{International Joint Conference}} on {{Natural Language Processing}} ({{EMNLP-IJCNLP}})},
  author = {Qian, Jing and Bethke, Anna and Liu, Yinyin and Belding, Elizabeth and Wang, William Yang},
  editor = {Inui, Kentaro and Jiang, Jing and Ng, Vincent and Wan, Xiaojun},
  year = {2019},
  month = nov,
  pages = {4755--4764},
  publisher = {Association for Computational Linguistics},
  address = {Hong Kong, China},
  doi = {10.18653/v1/D19-1482},
  urldate = {2025-06-10},
  langid = {american}
}

@techreport{ruthsCounterspeechTwitterField2016,
  title = {Counterspeech on {{Twitter}}: {{A Field Study}}},
  shorttitle = {Counterspeech on {{Twitter}}},
  author = {Ruths, Derek Ruths and Saleem, Haji Mohammed Saleem and Dillon, Kelly P. Dillon and Wright, Lucas Wright and Benesch, Susan Benesch},
  year = {2016},
  month = oct,
  address = {Washington, DC USA},
  institution = {Dangerous Speech Project},
  urldate = {2025-05-07},
  langid = {english}
}

@misc{sahaConsolidatingStrategiesCountering2024,
  title = {Consolidating {{Strategies}} for {{Countering Hate Speech Using Persuasive Dialogues}}},
  author = {Saha, Sougata and Srihari, Rohini},
  year = {2024},
  month = jan,
  number = {arXiv:2401.07810},
  eprint = {2401.07810},
  publisher = {arXiv},
  doi = {10.48550/arXiv.2401.07810},
  urldate = {2025-05-07},
  archiveprefix = {arXiv},
  langid = {american}
}

@inproceedings{sahaCounterGeDiControllableApproach2022,
  title = {{{CounterGeDi}}: {{A Controllable Approach}} to {{Generate Polite}}, {{Detoxified}} and {{Emotional Counterspeech}}},
  shorttitle = {{{CounterGeDi}}},
  booktitle = {Thirty-{{First International Joint Conference}} on {{Artificial Intelligence}}},
  author = {Saha, Punyajoy and Singh, Kanishk and Kumar, Adarsh and Mathew, Binny and Mukherjee, Animesh},
  year = {2022},
  month = jul,
  volume = {6},
  pages = {5157--5163},
  doi = {10.24963/ijcai.2022/716},
  urldate = {2025-04-25},
  langid = {english}
}

@misc{sahaCrowdCounterBenchmarkTypespecific2024,
  title = {{{CrowdCounter}}: {{A}} Benchmark Type-Specific Multi-Target Counterspeech Dataset},
  shorttitle = {{{CrowdCounter}}},
  author = {Saha, Punyajoy and Datta, Abhilash and Jana, Abhik and Mukherjee, Animesh},
  year = {2024},
  month = oct,
  number = {arXiv:2410.01400},
  eprint = {2410.01400},
  publisher = {arXiv},
  doi = {10.48550/arXiv.2410.01400},
  urldate = {2025-05-07},
  archiveprefix = {arXiv},
  langid = {american}
}

@article{sasseBreakingSilenceInvestigating2023,
  title = {Breaking the {{Silence}}: {{Investigating Which Types}} of {{Moderation Reduce Negative Effects}} of {{Sexist Social Media Content}}},
  shorttitle = {Breaking the {{Silence}}},
  author = {Sasse, Julia and Grossklags, Jens},
  year = {2023},
  month = oct,
  journal = {Proc. ACM Hum.-Comput. Interact.},
  volume = {7},
  number = {CSCW2},
  pages = {327:1--327:26},
  doi = {10.1145/3610176},
  urldate = {2025-05-16},
  langid = {american}
}

@inproceedings{saveskiStructureToxicConversations2021,
  title = {The {{Structure}} of {{Toxic Conversations}} on {{Twitter}}},
  booktitle = {Proceedings of the {{Web Conference}} 2021},
  author = {Saveski, Martin and Roy, Brandon and Roy, Deb},
  year = {2021},
  month = jun,
  series = {{{WWW}} '21},
  pages = {1086--1097},
  publisher = {Association for Computing Machinery},
  address = {New York, NY, USA},
  doi = {10.1145/3442381.3449861},
  urldate = {2025-04-17},
  isbn = {978-1-4503-8312-7},
  langid = {american}
}

@article{schiebGoverningHateSpeech,
  title={Governing hate speech by means of counterspeech on Facebook},
  author={Schieb, Carla and Preuss, Mike},
  booktitle={66th ica annual conference, at fukuoka, japan},
  pages={1--23},
  year={2016}
}

@article{schlugerProactiveModerationOnline2022,
  title = {Proactive {{Moderation}} of {{Online Discussions}}: {{Existing Practices}} and the {{Potential}} for {{Algorithmic Support}}},
  shorttitle = {Proactive {{Moderation}} of {{Online Discussions}}},
  author = {Schluger, Charlotte and Chang, Jonathan P. and {Danescu-Niculescu-Mizil}, Cristian and Levy, Karen},
  year = {2022},
  month = nov,
  journal = {Proc. ACM Hum.-Comput. Interact.},
  volume = {6},
  number = {CSCW2},
  pages = {370:1--370:27},
  doi = {10.1145/3555095},
  urldate = {2025-04-24},
  langid = {american}
}

@inproceedings{seeringShapingProAntiSocial2017,
  title = {Shaping {{Pro}} and {{Anti-Social Behavior}} on {{Twitch Through Moderation}} and {{Example-Setting}}},
  booktitle = {Proceedings of the 2017 {{ACM Conference}} on {{Computer Supported Cooperative Work}} and {{Social Computing}}},
  author = {Seering, Joseph and Kraut, Robert and Dabbish, Laura},
  year = {2017},
  month = feb,
  series = {{{CSCW}} '17},
  pages = {111--125},
  publisher = {Association for Computing Machinery},
  address = {New York, NY, USA},
  doi = {10.1145/2998181.2998277},
  urldate = {2025-05-13},
  isbn = {978-1-4503-4335-0},
  langid = {american}
}

@inproceedings{silvaFightingHateSpeech2023,
  title = {Fighting {{Against Hate Speech}}: {{A Case}} for {{Harnessing Interactive Digital Counter-Narratives}}},
  shorttitle = {Fighting {{Against Hate Speech}}},
  booktitle = {Interactive {{Storytelling}}},
  author = {Silva, Cl{\'a}udia},
  editor = {{Holloway-Attaway}, Lissa and Murray, John T.},
  year = {2023},
  pages = {159--174},
  publisher = {Springer Nature Switzerland},
  address = {Cham},
  doi = {10.1007/978-3-031-47655-6_10},
  isbn = {978-3-031-47655-6},
  langid = {english}
}

@article{sportelliLetsMakeDifference2025a,
  title = {``{{Let}}'s {{Make}} the {{Difference}}!'' {{Promoting Hate Counter-Speech}} in {{Adolescence Through Empathy}} and {{Digital Intergroup Contact}}},
  author = {Sportelli, Carmela and Cicirelli, Paolo Giovanni and Paciello, Marinella and Corbelli, Giuseppe and D'Errico, Francesca},
  year = {2025},
  journal = {Journal of Community \& Applied Social Psychology},
  volume = {35},
  number = {1},
  pages = {e70028},
  issn = {1099-1298},
  doi = {10.1002/casp.70028},
  urldate = {2025-05-07},
  copyright = {{\copyright} 2024 The Author(s). Journal of Community \& Applied Social Psychology published by John Wiley \& Sons Ltd.},
  langid = {english}
}

@inproceedings{suhAISocialGlue2021,
  title = {{{AI}} as {{Social Glue}}: {{Uncovering}} the {{Roles}} of {{Deep Generative AI}} during {{Social Music Composition}}},
  shorttitle = {{{AI}} as {{Social Glue}}},
  booktitle = {Proceedings of the 2021 {{CHI Conference}} on {{Human Factors}} in {{Computing Systems}}},
  author = {Suh, Minhyang (Mia) and Youngblom, Emily and Terry, Michael and Cai, Carrie J},
  year = {2021},
  month = may,
  series = {{{CHI}} '21},
  pages = {1--11},
  publisher = {Association for Computing Machinery},
  address = {New York, NY, USA},
  doi = {10.1145/3411764.3445219},
  urldate = {2024-08-07},
  isbn = {978-1-4503-8096-6}
}

@article{liDeModHolisticTool2025,
  title = {{{DeMod}}: {{A Holistic Tool}} with {{Explainable Detection}} and {{Personalized Modification}} for {{Toxicity Censorship}}},
  shorttitle = {{{DeMod}}},
  author = {Li, Yaqiong and Zhang, Peng and Gu, Hansu and Lu, Tun and Qiao, Siyuan and Shu, Yubo and Shao, Yiyang and Gu, Ning},
  year = {2025},
  month = may,
  journal = {Proc. ACM Hum.-Comput. Interact.},
  volume = {9},
  number = {2},
  pages = {CSCW061:1--CSCW061:24},
  doi = {10.1145/3710959},
  urldate = {2025-09-10},
  langid = {american}
}

@article{thomasGeneralInductiveApproach2003,
  title = {A General Inductive Approach for Qualitative Data Analysis},
  author = {Thomas, David R.},
  year = {2003},
  urldate = {2025-06-04},
  langid = {american}
}

@inproceedings{jungGreatChainAgents2022,
  title = {Great {{Chain}} of {{Agents}}: {{The Role}} of {{Metaphorical Representation}} of {{Agents}} in {{Conversational Crowdsourcing}}},
  shorttitle = {Great {{Chain}} of {{Agents}}},
  booktitle = {Proceedings of the 2022 {{CHI Conference}} on {{Human Factors}} in {{Computing Systems}}},
  author = {Jung, Ji-Youn and Qiu, Sihang and Bozzon, Alessandro and Gadiraju, Ujwal},
  year = {2022},
  month = apr,
  series = {{{CHI}} '22},
  pages = {1--22},
  publisher = {Association for Computing Machinery},
  address = {New York, NY, USA},
  doi = {10.1145/3491102.3517653},
  urldate = {2024-11-13},
  isbn = {978-1-4503-9157-3}
}

@article{biernackiSnowballSamplingProblems1981,
  title = {Snowball {{Sampling}}: {{Problems}} and {{Techniques}} of {{Chain Referral Sampling}}},
  shorttitle = {Snowball {{Sampling}}},
  author = {Biernacki, Patrick and Waldorf, Dan},
  year = 1981,
  month = nov,
  journal = {Sociological Methods \& Research},
  volume = {10},
  number = {2},
  pages = {141--163},
  issn = {0049-1241, 1552-8294},
  doi = {10.1177/004912418101000205},
  urldate = {2025-11-14},
  copyright = {https://journals.sagepub.com/page/policies/text-and-data-mining-license},
  langid = {english}
}

@phdthesis{cyprisEffectivenessCounterspeechMitigating2024,
  title = {The {{Effectiveness}} of {{Counterspeech}} in {{Mitigating Online Hate}}: {{Insights From}} a {{Multi-Method Investigation}}},
  shorttitle = {The {{Effectiveness}} of {{Counterspeech}} in {{Mitigating Online Hate}}},
  author = {Cypris, Niklas Felix},
  year = 2024,
  langid = {american},
  school = {Technische Universit\"at M\"unchen}
}

@inproceedings{difranzoUpstandingDesignBystander2018,
  title = {Upstanding by {{Design}}: {{Bystander Intervention}} in {{Cyberbullying}}},
  shorttitle = {Upstanding by {{Design}}},
  booktitle = {Proceedings of the 2018 {{CHI Conference}} on {{Human Factors}} in {{Computing Systems}}},
  author = {DiFranzo, Dominic and Taylor, Samuel Hardman and Kazerooni, Franccesca and Wherry, Olivia D. and Bazarova, Natalya N.},
  year = 2018,
  month = apr,
  series = {{{CHI}} '18},
  pages = {1--12},
  publisher = {Association for Computing Machinery},
  address = {New York, NY, USA},
  doi = {10.1145/3173574.3173785},
  urldate = {2025-11-21},
  isbn = {978-1-4503-5620-6},
  langid = {american}
}

@article{gennaroCounterspeechEncouragingUsers2025,
  title = {Counterspeech Encouraging Users to Adopt the Perspective of Minority Groups Reduces Hate Speech and Its Amplification on Social Media},
  author = {Gennaro, Gloria and Derksen, Laurenz and Abdelrahman, Aya and Broggini, Emma and Green, Mariya Alexandra and Haerter, Victoria Andrea and Heer, Elia and Heidler, Isabel and Kauer, Fiona and Kim, Han-Nuri and Landry, Benjamin and Levis, Alessio and Li, Jiazhen and {\c S}im{\c s}ir, {\c S}evval and Srbinovska, Iva and Vital, Robin Anna and Donnay, Karsten and Gilardi, Fabrizio and Hangartner, Dominik},
  year = 2025,
  month = jul,
  journal = {Scientific Reports},
  volume = {15},
  number = {1},
  pages = {22018},
  issn = {2045-2322},
  doi = {10.1038/s41598-025-05041-w},
  urldate = {2025-11-06},
  copyright = {2025 The Author(s)},
  langid = {english}
}

@article{golzarConvenienceSampling2022,
  title = {Convenience {{Sampling}}},
  author = {Golzar, Jawad and Noor, Shagofah and Tajik, Omid},
  year = 2022,
  month = dec,
  journal = {International Journal of Education \& Language Studies},
  volume = {1},
  number = {2},
  pages = {72--77},
  issn = {2767-4851},
  doi = {10.22034/ijels.2022.162981},
  urldate = {2025-11-14},
  langid = {english}
}

@inproceedings{jhaverBystandersOnlineModeration2024,
  title = {Bystanders of {{Online Moderation}}: {{Examining}} the {{Effects}} of {{Witnessing Post-Removal Explanations}}},
  shorttitle = {Bystanders of {{Online Moderation}}},
  booktitle = {Proceedings of the 2024 {{CHI Conference}} on {{Human Factors}} in {{Computing Systems}}},
  author = {Jhaver, Shagun and Rathi, Himanshu and Saha, Koustuv},
  year = 2024,
  month = may,
  series = {{{CHI}} '24},
  pages = {1--9},
  publisher = {Association for Computing Machinery},
  address = {New York, NY, USA},
  doi = {10.1145/3613904.3642204},
  urldate = {2025-11-05},
  isbn = {979-8-4007-0330-0},
  langid = {american}
}

@article{jhaverDoesTransparencyModeration2019,
  title = {Does {{Transparency}} in {{Moderation Really Matter}}? {{User Behavior After Content Removal Explanations}} on {{Reddit}}},
  shorttitle = {Does {{Transparency}} in {{Moderation Really Matter}}?},
  author = {Jhaver, Shagun and Bruckman, Amy and Gilbert, Eric},
  year = 2019,
  month = nov,
  journal = {Proc. ACM Hum.-Comput. Interact.},
  volume = {3},
  number = {CSCW},
  pages = {150:1--150:27},
  doi = {10.1145/3359252},
  urldate = {2025-11-28},
  langid = {american}
}

@incollection{ortloffSmallMediumLarge2025,
  title = {Small, {{Medium}}, {{Large}}? {{A Meta-Study}} of {{Effect Sizes}} at {{CHI}} to {{Aid Interpretation}} of {{Effect Sizes}} and {{Power Calculation}}},
  shorttitle = {Small, {{Medium}}, {{Large}}?},
  booktitle = {Proceedings of the 2025 {{CHI Conference}} on {{Human Factors}} in {{Computing Systems}}},
  author = {Ortloff, Anna-Marie and Martius, Florin and Meier, Mischa and Raimbault, Theo and Geierhaas, Lisa and Smith, Matthew},
  year = 2025,
  month = apr,
  series = {{{ACM Conferences}}},
  pages = {1--28},
  urldate = {2025-11-06},
  isbn = {979-8-4007-1394-1},
  langid = {american}
}

@inproceedings{sahaObserverEffectSocial2024,
  title = {Observer {{Effect}} in {{Social Media Use}}},
  booktitle = {Proceedings of the {{CHI Conference}} on {{Human Factors}} in {{Computing Systems}}},
  author = {Saha, Koustuv and Gupta, Pranshu and Mark, Gloria and Kiciman, Emre and De Choudhury, Munmun},
  year = 2024,
  month = may,
  pages = {1--20},
  publisher = {ACM},
  address = {Honolulu HI USA},
  doi = {10.1145/3613904.3642078},
  urldate = {2025-11-06},
  isbn = {979-8-4007-0330-0},
  langid = {english}
}

@article{sommetHowManyParticipants2023,
  title = {How {{Many Participants Do I Need}} to {{Test}} an {{Interaction}}? {{Conducting}} an {{Appropriate Power Analysis}} and {{Achieving Sufficient Power}} to {{Detect}} an {{Interaction}}},
  shorttitle = {How {{Many Participants Do I Need}} to {{Test}} an {{Interaction}}?},
  author = {Sommet, Nicolas and Weissman, David L. and Cheutin, Nicolas and Elliot, Andrew J.},
  year = 2023,
  month = jul,
  journal = {Advances in Methods and Practices in Psychological Science},
  volume = {6},
  number = {3},
  pages = {25152459231178728},
  issn = {2515-2459, 2515-2467},
  doi = {10.1177/25152459231178728},
  urldate = {2025-11-06},
  langid = {english}
}

@article{taylorAccountabilityEmpathyDesign2019,
  title = {Accountability and {{Empathy}} by {{Design}}: {{Encouraging Bystander Intervention}} to {{Cyberbullying}} on {{Social Media}}},
  shorttitle = {Accountability and {{Empathy}} by {{Design}}},
  author = {Taylor, Samuel Hardman and DiFranzo, Dominic and Choi, Yoon Hyung and Sannon, Shruti and Bazarova, Natalya N.},
  year = 2019,
  month = nov,
  journal = {Proc. ACM Hum.-Comput. Interact.},
  volume = {3},
  number = {CSCW},
  pages = {118:1--118:26},
  doi = {10.1145/3359220},
  urldate = {2025-11-28},
  langid = {american}
}

@misc{wangAdaptiveHumanAgentTeaming2025,
  title = {Adaptive {{Human-Agent Teaming}}: {{A Review}} of {{Empirical Studies}} from the {{Process Dynamics Perspective}}},
  shorttitle = {Adaptive {{Human-Agent Teaming}}},
  author = {Wang, Mengyao and Wu, Jiayun and Ma, Shuai and Li, Nuo and Zhang, Peng and Gu, Ning and Lu, Tun},
  year = 2025,
  month = apr,
  number = {arXiv:2504.10918},
  eprint = {2504.10918},
  primaryclass = {cs},
  publisher = {arXiv},
  doi = {10.48550/arXiv.2504.10918},
  urldate = {2025-06-11},
  archiveprefix = {arXiv},
  langid = {american}
}

@book{ullmannCounterspeechMultidisciplinaryPerspectives2024,
  title = {Counterspeech: {{Multidisciplinary Perspectives}} on {{Countering Dangerous Speech}}},
  shorttitle = {Counterspeech},
  editor = {Ullmann, Stefanie and Tomalin, Marcus},
  year = {2024},
  publisher = {Taylor \& Francis},
  urldate = {2025-05-07},
  langid = {american}
}

@misc{vidgenDetectingEastAsian2020,
  title = {Detecting {{East Asian Prejudice}} on {{Social Media}}},
  author = {Vidgen, Bertie and Botelho, Austin and Broniatowski, David and Guest, Ella and Hall, Matthew and Margetts, Helen and Tromble, Rebekah and Waseem, Zeerak and Hale, Scott},
  year = {2020},
  month = may,
  number = {arXiv:2005.03909},
  eprint = {2005.03909},
  publisher = {arXiv},
  doi = {10.48550/arXiv.2005.03909},
  urldate = {2025-05-14},
  archiveprefix = {arXiv},
  langid = {american}
}

@article{wachsEffectsPreventionProgram2023,
  title = {Effects of the {{Prevention Program}} ``{{HateLess}}. {{Together}} against {{Hatred}}'' on {{Adolescents}}' {{Empathy}}, {{Self-efficacy}}, and {{Countering Hate Speech}}},
  author = {Wachs, Sebastian and Krause, Norman and Wright, Michelle F. and {G{\'a}mez-Guadix}, Manuel},
  year = {2023},
  month = jun,
  journal = {Journal of Youth and Adolescence},
  volume = {52},
  number = {6},
  pages = {1115--1128},
  issn = {1573-6601},
  doi = {10.1007/s10964-023-01753-2},
  urldate = {2025-07-23},
  langid = {english}
}

@inproceedings{wilkFactbasedCounterNarrative2025,
  title = {Fact-Based {{Counter Narrative Generation}} to {{Combat Hate Speech}}},
  booktitle = {Proceedings of the {{ACM}} on {{Web Conference}} 2025},
  author = {Wilk, Brian and Shomee, Homaira Huda and Maity, Suman Kalyan and Medya, Sourav},
  year = {2025},
  month = apr,
  series = {{{WWW}} '25},
  pages = {3354--3365},
  publisher = {Association for Computing Machinery},
  address = {New York, NY, USA},
  doi = {10.1145/3696410.3714718},
  urldate = {2025-05-07},
  isbn = {979-8-4007-1274-6},
  langid = {american}
}

@misc{yuHateSpeechCounter2022,
  title = {Hate {{Speech}} and {{Counter Speech Detection}}: {{Conversational Context Does Matter}}},
  shorttitle = {Hate {{Speech}} and {{Counter Speech Detection}}},
  author = {Yu, Xinchen and Blanco, Eduardo and Hong, Lingzi},
  year = {2022},
  month = jun,
  number = {arXiv:2206.06423},
  eprint = {2206.06423},
  publisher = {arXiv},
  doi = {10.48550/arXiv.2206.06423},
  urldate = {2025-05-07},
  archiveprefix = {arXiv},
  langid = {american}
}

@misc{zhangConversationsGoneAwry2018a,
  title = {Conversations {{Gone Awry}}: {{Detecting Early Signs}} of {{Conversational Failure}}},
  shorttitle = {Conversations {{Gone Awry}}},
  author = {Zhang, Justine and Chang, Jonathan P. and {Danescu-Niculescu-Mizil}, Cristian and Dixon, Lucas and Hua, Yiqing and Thain, Nithum and Taraborelli, Dario},
  year = {2018},
  month = may,
  number = {arXiv:1805.05345},
  eprint = {1805.05345},
  publisher = {arXiv},
  doi = {10.48550/arXiv.1805.05345},
  urldate = {2025-05-07},
  archiveprefix = {arXiv},
  langid = {american}
}

@inproceedings{zhangYouCompleteMe2022a,
  title = {You {{Complete Me}}: {{Human-AI Teams}} and {{Complementary Expertise}}},
  shorttitle = {You {{Complete Me}}},
  booktitle = {Proceedings of the 2022 {{CHI Conference}} on {{Human Factors}} in {{Computing Systems}}},
  author = {Zhang, Qiaoning and Lee, Matthew L and Carter, Scott},
  year = {2022},
  month = apr,
  series = {{{CHI}} '22},
  pages = {1--28},
  publisher = {Association for Computing Machinery},
  address = {New York, NY, USA},
  doi = {10.1145/3491102.3517791},
  urldate = {2024-08-14},
  isbn = {978-1-4503-9157-3},
  langid = {american}
}

@article{zhaoMeasurePerceivedArgument2011,
  title = {A {{Measure}} of {{Perceived Argument Strength}}: {{Reliability}} and {{Validity}}},
  shorttitle = {A {{Measure}} of {{Perceived Argument Strength}}},
  author = {Zhao, Xiaoquan and Strasser, Andrew and Cappella, Joseph N. and Lerman, Caryn and Fishbein, Martin},
  year = {2011},
  month = mar,
  journal = {Communication Methods and Measures},
  volume = {5},
  number = {1},
  pages = {48--75},
  issn = {1931-2458, 1931-2466},
  doi = {10.1080/19312458.2010.547822},
  urldate = {2025-08-10},
  langid = {english}
}

@inproceedings{zhengWhatMakesGood2023b,
  title = {What {{Makes Good Counterspeech}}? {{A Comparison}} of {{Generation Approaches}} and {{Evaluation Metrics}}},
  shorttitle = {What {{Makes Good Counterspeech}}?},
  booktitle = {Proceedings of the 1st {{Workshop}} on {{CounterSpeech}} for {{Online Abuse}} ({{CS4OA}})},
  author = {Zheng, Yi and Ross, Bj{\"o}rn and Magdy, Walid},
  editor = {Chung, Yi-Ling and Bonaldi, Helena and Abercrombie, Gavin and Guerini, Marco},
  year = {2023},
  month = sep,
  pages = {62--71},
  publisher = {Association for Computational Linguistics},
  address = {Prague, Czechia},
  urldate = {2025-05-07},
  langid = {american}
}

@inproceedings{zhouIdentifyingSocialBias2022,
  title = {Towards {{Identifying Social Bias}} in {{Dialog Systems}}: {{Framework}}, {{Dataset}}, and {{Benchmark}}},
  shorttitle = {Towards {{Identifying Social Bias}} in {{Dialog Systems}}},
  booktitle = {Findings of the {{Association}} for {{Computational Linguistics}}: {{EMNLP}} 2022},
  author = {Zhou, Jingyan and Deng, Jiawen and Mi, Fei and Li, Yitong and Wang, Yasheng and Huang, Minlie and Jiang, Xin and Liu, Qun and Meng, Helen},
  editor = {Goldberg, Yoav and Kozareva, Zornitsa and Zhang, Yue},
  year = {2022},
  month = dec,
  pages = {3576--3591},
  publisher = {Association for Computational Linguistics},
  address = {Abu Dhabi, United Arab Emirates},
  doi = {10.18653/v1/2022.findings-emnlp.262},
  urldate = {2025-06-16},
  langid = {american}
}

@misc{zhuGeneratePruneSelect2021,
  title = {Generate, {{Prune}}, {{Select}}: {{A Pipeline}} for {{Counterspeech Generation}} against {{Online Hate Speech}}},
  shorttitle = {Generate, {{Prune}}, {{Select}}},
  author = {Zhu, Wanzheng and Bhat, Suma},
  year = {2021},
  month = jun,
  number = {arXiv:2106.01625},
  eprint = {2106.01625},
  publisher = {arXiv},
  doi = {10.48550/arXiv.2106.01625},
  urldate = {2025-05-07},
  archiveprefix = {arXiv},
  langid = {american}
}

\appendix 
\section{Appendix A: Questionnaire}
\label{app:questionnaire}
The final questionnaire assessed bystanders' responses on three dimensions: (1) perceived quality, (2) subjective acceptance, and (3) behavioural tendencies (pre–post). Items were translated from validated English scales and refined through pilot testing; Chinese items were back-translated to ensure conceptual equivalence. Table \ref{tab:questionnaire} lists the full bilingual items used in the study.

\begin{CJK*}{UTF8}{gbsn} 

\begin{table}[htbp]
\centering
\small
\caption{Questionnaire Items (Chinese/English)}
\label{tab:questionnaire}
\renewcommand{\arraystretch}{1.2}

\makebox[0.9\columnwidth][c]{%
\begin{tabular}{p{0.18\columnwidth} p{0.3\columnwidth}p{0.42\columnwidth}}
\toprule
\textbf{Dimension} & \textbf{Item (Chinese)} & \textbf{Item (English)} \\
\midrule
\textbf{Perceived Quality} &
文明罗伯特的反驳理由具有说服力。 &
The reason given in this Civilbot counterspeech is convincing. \\
&
文明罗伯特的反驳理由是强有力的。 &
The Civilbot counterspeech gives a strong reason. \\
\midrule
\textbf{Subjective Acceptance} &
文明罗伯特的反驳理由是可信的。 &
The reason given in this Civilbot counterspeech is believable. \\
&
文明罗伯特的反驳内容提出了我认为重要的理由。 &
The counterspeech by Civilbot provides a reason I consider important. \\
&
总体上，我同意文明罗伯特的反驳内容。 &
Overall, I agree with the Civilbot counterspeech. \\
\midrule
\textbf{Behavioural Tendencies (pre \& post)} &
我有信心应对这种仇恨言论。 &
I am confident that I can respond to such hate speech. \\
&
我想参与相关讨论。 &
I want to participate in the related discussion. \\
\bottomrule
\end{tabular}
}
\end{table}
\FloatBarrier
\end{CJK*}

\begin{CJK*}{UTF8}{gbsn}
\section*{Appendix B. Prompt Design}
\label{app:prompt}

\subsection*{B.1 Hate-Speech Screening Prompt}
\begin{table}[ht]
\centering
\small
\caption{Prompt used for the initial automatic screening of hate-speech candidates.}
\label{tab:prompt-hate}
\begin{tabular}{p{0.95\linewidth}}
\toprule
\textbf{Instruction}\\
\midrule
Definition of hate speech:  
\begin{itemize}
\item It is a weaponized statement  
\item Targeted at a specific social group  
\item Likely to cause emotional or psychological harm  
\item Includes hostility, dehumanization, or group contempt  
\end{itemize}

Return in JSON with keys:
\begin{itemize}
\item \texttt{label}: 1 (hate) or 0 (not hate)
\item \texttt{justification}: one-sentence explanation
\item \texttt{target\_group}: targeted group (e.g., women, LGBTQ, immigrants), or ``N/A''
\item \texttt{issue}: main topic (e.g., ``STEM vs humanities'')
\end{itemize}

Now evaluate:  
Context: ``\{q\}''  
Comment: ``\{a\}''
\\
\bottomrule
\end{tabular}
\end{table}
\FloatBarrier

\subsection*{B.2 Counterspeech Generation Prompt}
To generate counterspeech with balanced response lengths across rhetorical strategies, we adopted a \textit{soft-approximate approach}. Instead of imposing a strict character or token limit, we instructed GPT-5 to produce \emph{one short paragraph per speech act}, allowing mild natural variance in length while keeping outputs concise.
Under the model’s default sampling configuration, we applied conditional re-sampling only when outputs were clearly too long or too short, which maintained a moderate and balanced token-length distribution. Table~\ref{tab:prompt-counter} shows the exact prompt used to elicit the counterspeech responses.

\begin{table*}[t]
\centering
\small
\caption{Prompt for generating counterspeech sentences in Chinese with soft-approximate length control.}
\label{tab:prompt-counter}
\makebox[\textwidth][c]{%
\begin{tabular}{p{0.9\textwidth}}
\toprule
\textbf{Instruction}\\
\midrule
You are a counterspeech generation expert for online hate speech in Chinese.  
Task: Generate exactly \textbf{ONE short paragraph} (one speech act) in Chinese matching 100\% the requested rhetorical dimensions.  
If any rule is broken, regenerate.  
Keep the response concise but allow natural length variation to support soft-approximate balance across strategies.

\textbf{RULES}
\begin{enumerate}
\item Sentence Type  
  \begin{itemize}
  \item Q: must be a question only, ending with ``？'' or Chinese question particles; no statements.  
  \item Non-Q: must be a declarative sentence only; no question marks or question words.  
  \end{itemize}

\item Tone  
  \begin{itemize}
  \item Positive: friendly, cooperative, supportive.  
  \item Negative: sarcastic, mocking, critical, emotionally intense.  
  \end{itemize}

\item Strategy Intent  
  \begin{itemize}
  \item Cognitive: 
    \begin{itemize}
      \item $\triangleright$ Rebut falsehoods (e.g., highlight hypocrisy, logical flaws, unreliable sources),
      \item $\triangleright$ Highlight truth (e.g., provide facts, suggest proper action, warn of consequences).
    \end{itemize}
  \item Affective: 
    \begin{itemize}
      \item $\triangleright$ Denounce perpetrators (e.g., explicitly identify hate, evoke shame, raise alarm),
      \item $\triangleright$ Support targets (e.g., express empathy, solidarity, or emotional validation).
    \end{itemize}
  \end{itemize}
\end{enumerate}

Input Variables:\\
Type: \{stype\} / \{tone\} / \{intent\}\\
Hatespeech: \{hate\_speech\}\\
Counterspeech:\\
\bottomrule
\end{tabular}
}
\end{table*}

\end{CJK*}

\end{document}